\documentclass[reprint,prl,showpacs,superscriptaddress]{revtex4-1}

\usepackage{amssymb}
\usepackage{amsmath}    
\usepackage{graphicx}   
\usepackage{verbatim}

\usepackage{color}      

\begin{document}
	
	\title{Scaling laws for direct laser acceleration in a radiation-reaction dominated regime}
	
	\author{M. Jirka}
	\affiliation{Institute of Physics of the Czech
	Academy of Sciences, ELI Beamlines, Na Slovance
		2, 182 21 Prague, Czech Republic}
	\affiliation{Faculty of Nuclear Sciences and Physical Engineering, Czech
		Technical University in Prague, Brehova 7, 115 19 Prague, Czech Republic}
	
	\author{M. Vranic}
	\affiliation{GoLP/Instituto de Plasmas e Fus\~{a}o Nuclear, Instituto Superior 
		T\'{e}cnico, Universidade de Lisboa, Lisbon, 1049-001, Portugal}
	
	\author{T. Grismayer}
	\affiliation{GoLP/Instituto de Plasmas e Fus\~{a}o Nuclear, Instituto Superior 
		T\'{e}cnico, Universidade de Lisboa, Lisbon, 1049-001, Portugal}
	
	\author{L. O. Silva}
	\affiliation{GoLP/Instituto de Plasmas e Fus\~{a}o Nuclear, Instituto Superior 
		T\'{e}cnico, Universidade de Lisboa, Lisbon, 1049-001, Portugal}

	\begin{abstract}
		We study electron acceleration within a sub-critical plasma channel irradiated by an ultra-intense laser pulse ($a_0>1
		00$ or $I>10^{22}~\mathrm{W/cm^2}$).
		In this regime, radiation reaction significantly alters the electron dynamics.
		This has an effect not only on the maximum attainable electron energy but also on the phase-matching process between betatron motion and electron oscillations in the laser field.
		Our study encompasses analytical description, test-particle calculations and 2-dimensional particle-in-cell simulations. 
		We show single-stage electron acceleration
		to multi-GeV energies within a 0.5 mm-long channel 
		and provide guidelines how to obtain 
		energies beyond 10 GeV using optimal initial configurations.
		We present the required conditions in a form of explicit analytical scaling laws that 
		can be applied to plan the future electron acceleration experiments. 
	\end{abstract}
	
	
	\maketitle
	
	In the interaction of under-dense plasma with a strong laser pulse, plasma 
	electrons 
	can be accelerated to relativistic velocities 
	\cite{Tajima1979}. 
	Depending on the laser pulse duration, different mechanisms are responsible for 
	electron acceleration.
	Using laser-wakefield acceleration (LWFA) tecnique, one can obtain quasi-monenergetic electron energy
	distribution \cite{LWFA_mono_1, LWFA_mono_2, LWFA_mono_3, Leemans2014}.
	The highest electron energy achieved in experiments with LWFA to-date is 7.8 GeV \cite{Gonsalves2019}. 
	For an efficient LWFA, the laser pulse duration $\tau_0$ should be on the order of $1/\omega_p$, 
	where $ \omega_{\mathrm{p}} $ is the electron plasma frequency. Short lasers ($\tau_0\sim 30~ \mathrm{fs}$) are 
	required to achieve this.  
	
	For a long laser pulse $\tau_0\gg 1/\omega_{\mathrm{p}} $, where LWFA is not efficient,
	an alternative acceleration mechanism can be explored. 
	Upon laser propagation through the underdense plasma, the ponderomotive force 
	expels electrons sideways resulting in the formation of a positively charged 
	plasma channel.
	The formation of longitudinal plasma wave used for particle acceleration in LWFA is prevented by the length of the laser pulse. 
	However, the formation of long-range transverse fields is still possible, due to the charge displacement and the electron current. 
	Electrons propagating at relativistic velocities aligned with the channel axis undergo betatron oscillations, similarly as in LWFA bubble regime.
	At the same time, they are also experiencing the transverse laser fields. 
	If the frequency of betatron oscillations is close to the Doppler-shifted 
	laser frequency, then the laser energy can be 
	efficiently coupled to the electron (so-called betatron resonance).
	This regime of electron acceleration is referred to as the direct 
	laser 
	acceleration (DLA) \cite{Pukhov1999}.
	The rich physics of this setup has raised a keen interest among researchers in the past few years
	\cite{Tsakiris2000,Qiao2005Electron,Qiao2005Quasistatic,Li2011,Gu2011,Liu2013,Shaw2014,Robinson2015,Hu2015,
		Arefiev2016Beyond,Khudik2016,Arefiev2016Spontaneous,Zhang2016,Huang2016,Huang2017Relativistic,Jiang2018,Geng2018,Wang2018}. 
	It was reported that the efficiency of DLA may be enhanced using various strategies: parametric 
	amplification of betatron oscillations \cite{Arefiev2012,Arefiev2014,Arefiev2015,Huang2017Nonlinear}, applying 
	additional longitudinal electric field \cite{Robinson2013, Khudik2018Far} or by "breaking of the adiabaticity that precludes electron 
	energy retention" \cite{Robinson2017}.
	Even a few DLA experiments have been performed using weakly relativistic laser intensities \cite{Gahn1999,Mangles2005}. 

	Ultra-intense laser beams will become available through the next generation of 10~PW-class laser systems \cite{Weber2017,Gales2016,Zou2015}.
	Interaction with a very intense pulse can provide a strong acceleration through DLA, but the underlying physics becomes even more complex than before. 
	Relativistic electrons in a strong electromagnetic field 
	lose energy by radiation, which in turn alters their trajectories.  
	\cite{Landau1975}.
	This phenomenon is called radiation reaction (RR) and is expected to affect electron motions at laser intensities exceeding $ 10^{22}~\mathrm{W/cm^{2}} $.  
	A small local change in electron trajectory can make a difference as to whether the betatron 
	resonance is achieved by that electron or not
	\cite{Huang2017Nonlinear}. 
	It is therefore necessary to assess the effect of radiation reaction on DLA 
	for the lasers of ultra-high intensities, i.e. when
	$ 
	a_{0}\gtrsim100 $,
	where $ 
	a_{0} $ is the dimensionless 
	normalized vector potential defined by $a_0\simeq 86~ \sqrt{I_0[10^{22}\mathrm{W/cm^2}]}~ \lambda_{0}[\mu\mathrm{m}] $, with $I_0$ and $\lambda_{0}$ representing the laser intensity and wavelength respectively. 
	%
	%
	%
	
	%
	The question regarding the role of radiation reaction in DLA was addressed in several recent 
	works.
	The studies encompass the effect of radiation reaction on the formation of an 
	ultradense 
	helical 
	electron bunch \cite{Qiao2017, Chang2017} and emission of $\gamma$-rays
	\cite{Huang2017Collimated, Wang2019}.
	%
	%
	Somewhat counter-intuitively, radiation damping can be favourable for particle acceleration. 
	One way to contribute is through allowing particle trapping 
	near 
	the channel axis in situations where all electrons would be expelled if no radiation 
	was emitted. The trapped electrons interact with the peak laser intensity and can potentially
	achieve multi-GeV energies \cite{Ji_Pukhov_trapping_2014, Vranic2018, Liu2018}.  
	%
	%
	Another way to contribute is through the change of resonant conditions.  
	Even though radiation reaction may reduce the global maximum achievable energy of accelerated 
	electrons, it can strongly enhance acceleration of some electrons \cite{Gong2019}.

	Obtaining a precise description of DLA is a challenge even without considering radiation reaction.
	By extending the mathematical tools previously developed for classical relativistic DLA at moderate 
	laser intensities ($a_0\lesssim10$) \cite{Khudik2016}, this paper aims to pave the path for a quantitative description of DLA including radiation reaction.
	%
	%
	%
	In particular, we explore how RR changes the onset 
	of betatron resonance and provide analytical predictions for the final electron energy. 
	We discuss different aspects that affect or limit the acceleration, and find  
	the promising parameter ranges and scaling laws to guide future experiments. 
	The laws are verified with a comprehensive numerical study, involving test particles in ideal conditions, as well as full-scale self-consistent 2D particle-in-cell simulations.
	The setup considered in this work is a pre-formed plasma channel  irradiated by an intense laser 
	pulse that accelerates plasma electrons which are initially at rest (somewhere within the channel or at the channel 
	walls).
	Our findings show that a fraction of particles that can achieve the betatron resonance is greatly increased by RR, which
	allows for high charge content  in the accelerated electron beam.
	We quantify which are the necessary initial conditions and expected asymptotic energies for those particles. 

	This manuscript is organized as follows. In section I, we 
	provide the analytical description of electron dynamics within a plasma channel 
	in the 
	presence of an intense laser field.
	We consider cases with and without radiation reaction in a simplified field configuration.
	In section II, test particle simulations are presented,
    the obtained data are compared with the analytical predictions, and the validity of used approximations is verified.
   Section III features 2D Particle-In-Cell (PIC) simulations with a pre-formed plasma channel,
	while the summary of the work is given in section IV.

	\section{I. Electron dynamics in a plasma channel 
		with a laser field}
	In this section, we analyze the acceleration of a single electron in a symmetric 
	plasma channel irradiated by an ultra-intense linearly 
	polarized laser.
	%
	Using DLA scheme, the electron can be efficiently accelerated only if it achieves
	the betatron resonance.
	Whether this condition is fullfilled depends on initial conditions of the electron 
	(e.g. distance from the axis, initial momentum, etc.) and the background plasma density. 
	In addition, as the electron performs betatron oscillations in a strong 
	electromagnetic background field, it loses its energy by emitting 
	photons which, in turn, alters its dynamics \cite{Landau1975}.
	All these effects have an impact on achieving the
	betatron resonance, and thus on the overall dynamics of the electrons within the plasma channel.

	We first introduce our configuration and an analytical description of a particle 
	performing the coupled oscillations. 
	This motion has an invariant $\mathcal{I}$ that can be used to obtain the resonance condition. 
	The next step is to estimate how much energy particles can gain over a 
	certain acceleration distance, still without radiation reaction. 
	We show later on that the principal effect of radiation reaction is to limit to the maximum energy achievable on a resonant trajectory. More specifically, radiation reaction breaks the invariant $\mathcal{I}$ and it is possible to quantify this change over a resonant cycle. 
	A decrease in $\mathcal{I}$ results in a gradual change of the resonant condition over time, which allows
	the electron to become resonant even if its initial conditions were far from optimal. 
	%
	
	\subsection{Integral of motion in simplified electromagnetic configuration}
	
\begin{figure}
		\centering
		\includegraphics[width=1\linewidth]{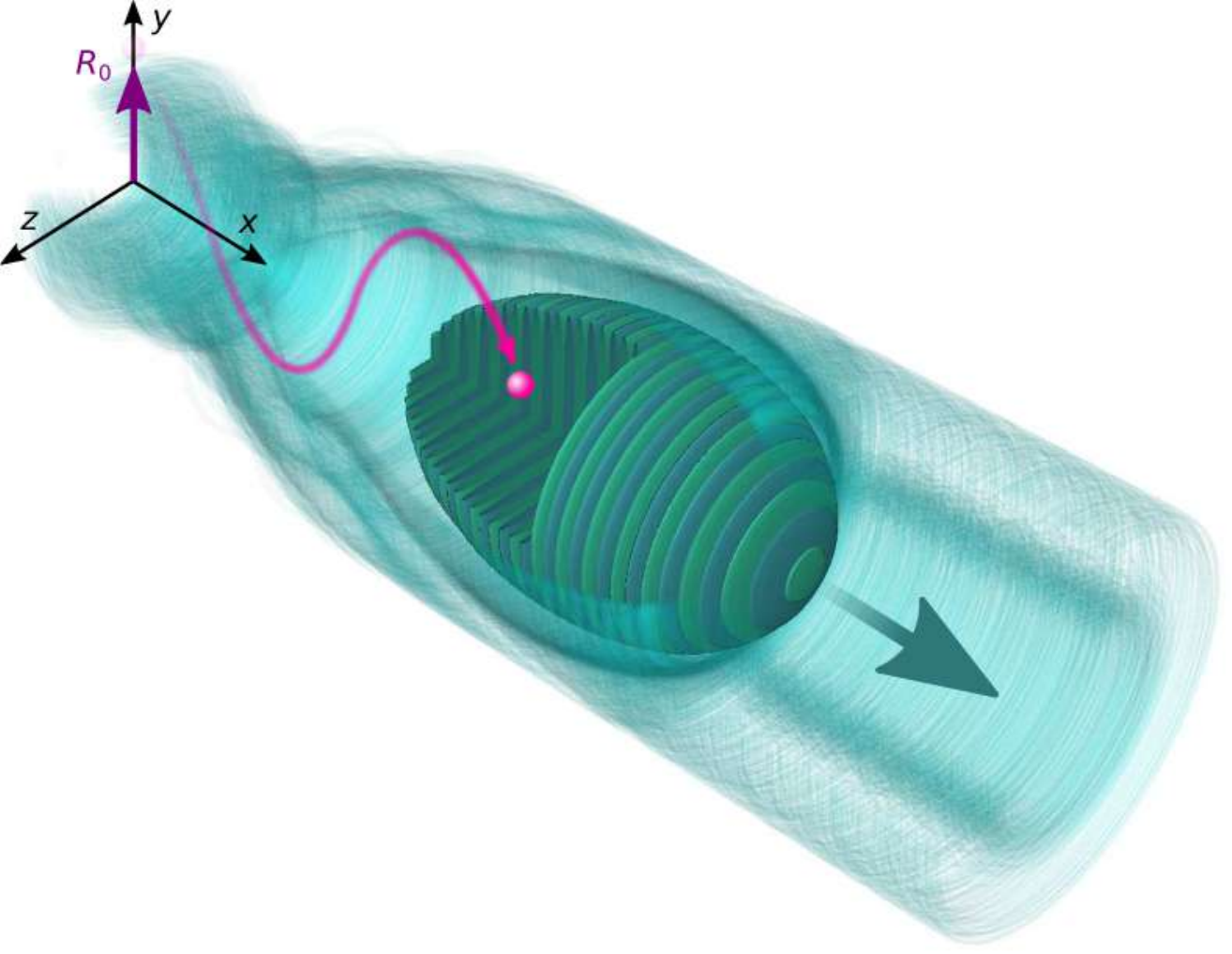}
		\caption{Simulation setup: an intense laser pulse accelerates an 
			electron in a
			cylindrically symmetric plasma channel.}
		\label{fig:01}
	\end{figure}

	A rigorous analytical description of our setup requires a few simplifications. 
	Similarly as in previous works, in this section we consider the laser to be a plane wave, and the channel fields
	to be a linear function of the distance from the propagation axis $x$ \cite{Arefiev2014, Khudik2016, Arefiev2016Spontaneous}. 
	Consequently, the channel fields are radially symmetric with respect to the $ x $-axis.
	The electron can be placed initially at different 
	radial positions 
	inside the plasma channel or on channel walls (see figure~\ref{fig:01}).
	We consider electrons starting at rest, and in the $(x,y)$ plane (the plane defined by the laser polarisation and propagation direction). 
	The laser propagates in positive $x$ direction and the fields are given by $\mathbf{E}_\mathrm{L}=E_{0}\sin\phi~ \mathbf{\hat{y}}$,
	$\mathbf{B}_\mathrm{L}=B_{0}\sin\phi~ \mathbf{\hat{z}}$,
	%
	where $ E_{0}$ and $B_0 $ are the amplitudes of the electric and magnetic field 
	and $ \phi $ is 
	the phase of the wave.
	The phase velocity of the laser is assumed to be equal to the speed of 
	light $ c $, which is justified by the low plasma density and high laser intensity (this is verified in section II).
	%

	The electromagnetic field experienced by the electron is the combination of the 
	laser field and the fields emerging due to the 
	displacement of plasma electrons in 
	the  
	channel (channel fields) \cite{Pukhov1999}.
	These self-generated quasi-static channel fields are the
	radial electric field and the azimuthal magnetic field 
	\begin{equation}\label{channel_fields_eq}
	\mathbf{E}_\mathrm{C}=f\dfrac{m_{e}\omega_{\mathrm{p}}^{2}}{2e}\mathbf{r},\quad 
	\mathbf{B}_\mathrm{C}=(1-f)\dfrac{m_{e}\omega_{\mathrm{p}}^{2}}{2ec}\mathbf{r}\times\mathbf{v},
	\end{equation}
	where $\mathbf{r}=y~\mathbf{\hat{y}}+z~\mathbf{\hat{z}}$ is perpendicular to the channel axis and $\mathbf{v}=v~ \mathbf{\hat{x}}$ is the velocity of the flow. The numerical factor $f$ depends on the fraction of electrons within the plasma channel and takes values between $ 0\leq f\leq1 $  \cite{Khudik2016,Arefiev2016Spontaneous,Huang2016,Arefiev2012,Huang2017Nonlinear,Arefiev2016Beyond}. 
	The transversely expelled electrons generate the radial electric field, while electrons accelerated forward within the channel form a current that generates the azimuthal magnetic field.
	Usually, the higher the background plasma density, the lower the value of $f$ \cite{Vranic2018}.
	%
	%
	%
	%
	In other words, the channel fields are linearly dependent on a radial distance 
	from the channel axis, but the electric field $E_\mathrm{C}$ and the magnetic field $B_\mathrm{C}$ do not 
	necessarily have the same magnitude.
	The total electromagnetic field experienced by the electron is then given by  
	\begin{equation}\label{E_total}
	\mathbf{E}=\mathbf{E}_\mathrm{L} + \mathbf{E}_\mathrm{C}, \quad 
	\mathbf{B}=\mathbf{B}_\mathrm{L} + \mathbf{B}_\mathrm{C}.
	\end{equation}
	%
	%
	%
	%
	%
	The field structure defined in equation \eqref{E_total} induces electron oscillations due to the laser field as well as betatron 
	oscillations at the same time.
	The background plasma density $n_{\mathrm{p}}$ affects 
	the electron motion 
	since the magnitude of the self-generated channel 
	fields is proportional to the plasma frequency: 
	$\omega_\mathrm{p}^{2} = 4\pi e^2 n_{\mathrm{p}} /m_e$.
	We, therefore, expect DLA to be sensitive to the initial conditions of 
	the electron, the intensity of the laser pulse and the density within the plasma channel.
	
	Without radiation reaction, the electron motion in the channel is only 
	governed by the Lorentz force:
	\begin{equation}\label{EoM}
	\dfrac{d \mathbf{p}}{dt}=-e\left( 
	\mathbf{E}+\frac{\mathbf{p}}{\gamma m_e c} \times\mathbf{B}\right),
	\end{equation}
	where $\gamma=\sqrt{1+(|\mathbf{p}|/m_{e}c)^2}$ is the relativistic Lorentz 
	factor and $\mathbf{p}$ is the electron momentum. 
	For particles propagating in the positive $ x $-direction, both $\mathbf{E}_\mathrm{C}$ and $\mathbf{B}_\mathrm{C}$
	contribute in a similar manner: they provide a restoring force that pushes electrons towards the channel axis.  
	In fact, the value of the numerical factor $f$ from equation 
	\eqref{channel_fields_eq} is not important because the 
	restoring 
	force is actually proportional to $| \mathbf{E}_\mathrm{C}|+ | \mathbf{B}_\mathrm{C}|$.
	
	%
	From the Hamiltonian of the electron, one obtains an integral of motion $\mathcal{I}$. For a particle that is 
	initially in the ($x,y$) plane with $z=0$, the integral
	of motion can be written as \cite{Khudik2016}
	\begin{equation}\label{IOM}
	\mathcal{I}=\gamma-\dfrac{p_{x}}{m_{e}c}+
	\dfrac{\omega_{\mathrm{p}}^{2}y^{2}}{4c^{2}},
	\end{equation}
	where
	$ p_{x} $ is the component of the electron 
	momentum in the direction of wave propagation.
	For a better intuitive understanding, it is useful to note 
	that the first two terms of the integral of motion are the same 
	as for the  particle interacting with a plane wave  in vacuum, while the third term
	accounts for the transverse harmonic oscillations within the channel.  
	The value of $ \mathcal{I} $ defines a limit on the energy attainable in the 
	system if all dissipation mechanisms are neglected.
	For example, in the case of the electron initially at rest 
	placed on the channel axis ($y=0$), the value of the integral 
	of motion is $ \mathcal{I}=1 $ 
	\cite{Khudik2016}.

	\subsection{Betatron resonance}
	
	%
	Due to the interaction with the channel fields, the electrons perform betatron 
	oscillations with the frequency given by 
	$\omega_{\beta}={\omega_{\mathrm{p}}/\sqrt{2\gamma}}$ \cite{Pukhov2002}. 
	We define the betatron period to be $ T_{\beta}=2\pi/\omega_{\beta} $.
	%
	For electrons propagating in the same direction as the laser, the frequency of 
	the quiver motion due to the laser field depends on how fast the electron moves in the $ x $-direction. 
	If the velocity of the electron is $ v_{x} $ while the laser 
	propagates at approximately the speed of light $c$ (underdense plasma)
	the Doppler-shifted laser frequency can be expressed as
	$\omega_{\mathrm{L}} =\omega_{0}\left(  1-{v_{x}}/{c}\right)$.
	%
	The electron in the channel can achieve an efficient acceleration if the conditions for the betatron resonance are met, i.e. when the Doppler-shifted laser frequency $ 
	\omega_{\mathrm{L}} $
	is close to the betatron frequency \cite{Pukhov1999}
	\begin{equation}\label{resonance}
	\omega_{\mathrm{L}}\simeq \omega_{\beta}.
	\end{equation} 
	%
	If we assume that the momentum of an
	ultra-relativistic electron accelerated in the channel is predominantly pointing
	in the forward direction ($ p_{x}\gg p_{y}\gg m_{e}c $), then 
	%
	the forward momentum can be expressed \cite{Khudik2016} as $p_{x}\simeq\gamma m_{e}c[1-(1/2)\left(p_{y}/\gamma m_e c\right)^2 ].$
	%
	After replacing $p_x$ into equation \eqref{IOM}, we obtain 
	\cite{Wang2018}
	\begin{equation}\label{IOP_Wang}
	\mathcal{I}\simeq\frac{1}{2\gamma} \left( \dfrac{p_{y}}{ m_e c}\right)^2+
	\dfrac{\omega_{\mathrm{p}}^{2}R^{2}}{4c^{2}}.
	\end{equation}
	%
	Equation~\eqref{IOP_Wang} allows to obtain an expression for 
	$p_{y}\simeq m_e c\sqrt{2\gamma\mathcal{I}}\cos \psi$, where the amplitude of the oscillations is written as $y=R=R_0 \sin\psi$,
	and $ \psi $ is the betatron phase.
	%
	%
	%
	The Doppler-shifted laser frequency is now
	\begin{equation}\label{Dopp_wang}
	\omega_{\mathrm{L}}\simeq\omega_{0}
	\dfrac{\mathcal{I}}{\gamma}\cos^{2} \psi.
	\end{equation}
	%
	%
	The resonant condition $\omega_L\simeq \omega_{\beta}$ applies to an average over the entire cycle, and not for the instantaneous values of $\psi$. For the condition to be satisfied, the required particle energy is 
	\begin{equation}\label{gamma_res}
	\gamma^{*}\approx2\mathcal{I}^{2}\left( 
	\dfrac{\omega_{0}}{\omega_{\mathrm{p}}}\right)^{2}\cos^{4}\psi.
	\end{equation}
	$\mathcal{I}$ being an invariant, we can also write
	$\mathcal{I}=\mathcal{I}_0$ and $\mathcal{I}_0=1+\left[ \omega_\mathrm{p} 
	R_0/(2c)\right] ^2$ for a particle initially at rest. This means that the 
	maximum $\gamma^{*}_\mathrm{max}$ and average $\left\langle \gamma^{*} 
	\right\rangle$ in a resonant cycle can be expressed as
	\begin{equation}\label{gamma_final_max}
	\gamma^{*}_{\mathrm{max}} \simeq 2\mathcal{I} ^{2}
	\left( \dfrac{\omega_{0}}{\omega_{\mathrm{p}}}\right) ^{2}, \quad \left\langle 
	\gamma^{*} \right\rangle \simeq\dfrac{3}{4}\mathcal{I} ^{2}
	\left( \dfrac{\omega_{0}}{\omega_{\mathrm{p}}}\right) ^{2},
	\end{equation}
	using cycle-averaged value $\left\langle \cos^{4}\psi \right\rangle  = 3/8$.
	According to equation \eqref{gamma_final_max}, the resonant electron energy does not depend on the 
	laser intensity but solely on the initial position and 
	the background plasma density.
	However, the intensity will determine how long it may take an electron to get accelerated towards the resonant energy.
	In the next subsection, we will discuss the work of the electromagnetic field 
	for optimal particle acceleration, with a special consideration of resonant 
	electrons. 
	
	\subsection{Optimal acceleration for quasi-resonant electrons}

	%
	%
	%
	%
	%
	%
	
\begin{figure}
		\centering
		\includegraphics[width=1.0\linewidth]{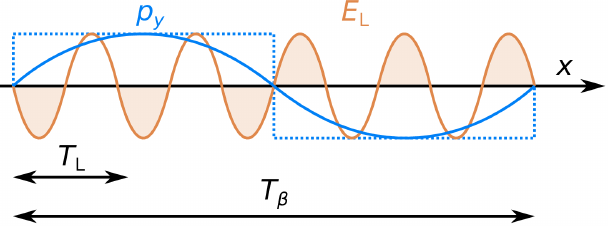}
		\caption{Shaded regions illustrate constructive work of the laser field $ 
			E_{\mathrm{L}} $ acting on the electron with $ p_{y} $ within one 
			betatron 
			cycle $ T_{\beta} $. $ T_{\mathrm{L}} $ is the laser period. The dotted line 
			represents the approximation of $ \sin\psi $ as a 
			step function. }
		\label{fig:02}
	\end{figure}

	The work performed by the electric field of the laser on one particle is given by $dW=-e\int\mathbf{E}\cdot \mathbf{p}/(\gamma m_e) ~dt$. 
	In our configuration, the fields are purely transverse and only the $p_y$ component is relevant
	\begin{equation}\label{work_of_field}
	\dfrac{d W}{d t} 
	\simeq    -eE_{0}c\sqrt{\frac{2\mathcal{I}}{\gamma}}\sin\phi\cos\psi.
	\end{equation}
	This expression is composed of two oscillatory functions. It is therefore crucial to determine under which conditions the averaged work does not vanish. Intuitively, the initial phase of the particle should be favourable and the oscillation frequencies should be of the same order.
	This can be understood by looking at figure~\ref{fig:02}. 
	The field is doing constructive work when $\mathbf{p}_y$ is anti-parallel with 
	$\mathbf{E_L}$. 
	If $T_\mathrm{L}\ll T_\beta$, then during one betatron oscillation, there are 
	many oscillations of the laser field whose work in the positive and negative 
	half-cycles would cancel making the total work vanish.   
	By approximating the slowly varying function $\sin\psi$ as a step 
	function, then at best we can obtain one entire 
	laser cycle $T_\mathrm{L}$ of the constructive work within one $T_\beta$ (sections
	of the cycle with constructive contribution are shaded in figure \ref{fig:02}).
	This would give $\left\langle\sin\phi\cos\psi \right\rangle 
	_{\mathrm{max}}\simeq \left\langle|\sin\phi |\right\rangle 
	T_\mathrm{L}/T_\beta\simeq (2/\pi)T_\mathrm{L}/T_\beta$.
	If we apply the same idea, but account on average for the fact that betatron 
	oscillations are not a step function, we get $\left\langle | \sin\phi\cos\psi | \right\rangle 
	_{\mathrm{max}}\simeq (2/\pi)^2T_\mathrm{L}/T_\beta$. 
	The work of the field is therefore expected to be less efficient when there is 
	a large discrepancy between the oscillation periods $T_\mathrm{L}$ and $T_\beta$. 
	Inserting our estimates into equation \eqref{work_of_field} and expressing the values of $T_\mathrm{L}$ and $T_\beta$ as a function of  $\mathcal{I}$, $\omega_0 $, and $\omega_{\mathrm {p}}$, we can estimate the 
	maximum attainable constructive power
	\begin{equation}\label{Off_resonance_work}
	\left\langle \left| \dfrac{d W}{d t}\right|\right\rangle 
	\simeq 
	\frac{8}{\pi^2} 
	~\dfrac{eE_{0}c}{\sqrt{\mathcal{I}}}\dfrac{\omega_{\mathrm{p}}}{\omega_{0}}.
	\end{equation}
	\begin{figure}[t!]
		\centering
		\includegraphics[width=8.5cm]{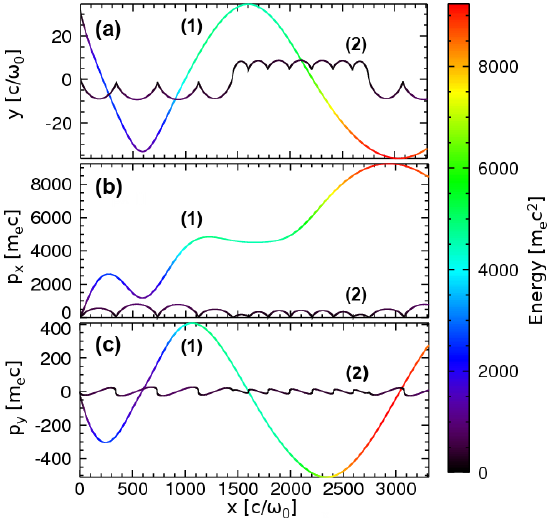}
		\caption{\label{fig:05}
			(a) Electron trajectory, (b) longitudinal and (c) transverse momenta during the 
			acceleration within a plasma channel where $ \omega_{\mathrm{p}}/\omega_{0}=0.5 $
			and for two 
			different initial 
			electron positions: (1) $ R_{0}=28\,c/\omega_{0} $ and (2) $ 
			R_{0}=0$. In both cases, the laser 
			pulse
			has $ 
			a_{0}=100 $. The color bar 
			represents the 
			relativistic $ \gamma $ factor of electrons.
		}
	\end{figure}
	
	This represents the power for the best case scenario --- there is no guarantee that any particles would meet this condition. 
	Figure \ref{fig:05} illustrates a resonant vs. non-resonant particle trajectory. 
	It is clear that some particles never get accelerated, despite being in the strong field. 
	A strict calculation would be possible for resonant particles for which  
	$\left\langle\sin\phi\cos\psi \right\rangle 
	_{\mathrm{max}}\simeq 1/2$.  
	This yields a solution similar to what equation \eqref{Off_resonance_work} predicts for $T_\mathrm{L}\simeq T_\beta$.
	
	Equation \eqref{Off_resonance_work} considers the local amplitude of the electric field $E_0$. 
	As we are usually dealing with laser pulses, this local field will vary in space and time. 
	We could assume, for example, a $\sin^2$ shape function for the laser temporal envelope. 
	This would lower the total work done by the laser from the beginning to the time when the particle gets to the highest field region approximately by a factor of two.  
	From that we can derive the expected cutoff energy versus time 
	\begin{equation}\label{gamma_t}
	\mathcal{E}[\mathrm{MeV}] \simeq 0.38 ~
	\dfrac{a_{0}}{\sqrt{\mathcal{I}}}~\dfrac{\omega_{\mathrm{p}}}{\omega_{0}} 
	\frac{t[\mathrm{fs}]}{\lambda_0[\mu\mathrm{m}]}
	\end{equation}
	which can also be presented as a function of the acceleration distance $l_\mathrm{acc}$
	\begin{equation}\label{gamma_l}
	\mathcal{E}[\mathrm{MeV}] \simeq 1.27 ~
	\dfrac{a_{0}}{\sqrt{\mathcal{I}}}~\dfrac{\omega_{\mathrm{p}}}{\omega_{0}} 
	\frac{l_\mathrm{acc}}{\lambda_0}.
	\end{equation}
	From \eqref{gamma_t} and \eqref{gamma_l}, we observe that particles which start with a low $\mathcal{I}$ can experience a more efficient acceleration.
	Moreover, their resonant $\gamma$ is lower according to equation 
	\eqref{gamma_final_max}, so the resonance may be easier to achieve. 
	Consequently, for any setup with a predetermined acceleration distance, we may have particles at a preferential initial $\mathcal{I}$ to achieve the betatron resonance at a highest value of $\gamma$. We can then extract the value for the optimum initial distance from the channel axis
	
	\begin{equation}\label{optimal_r}
	\frac{R_\mathrm{opt}}{\lambda_0}=0.4~\left(a_{0}\dfrac{l_\mathrm{acc}}{\lambda_0} \right) ^{1/5}\left( 
	\dfrac{\omega_{0}}{\omega_{\mathrm{p}}}\right) ^{2/5}. 
	\end{equation}
	The particles that were initially at the distance $R_\mathrm{opt}$ as defined by equation \eqref{optimal_r} have the best conditions for achieving the 
	betatron resonance and getting accelerated. 
	This remains true as long as the interaction is without radiation losses, i.e. the integral of motion is unchanged $\mathcal{I}\simeq\mathcal{I}_0$.
	We draw attention to the reader that the above calculations did not account for 
	a transverse shape of the laser pulse, as the derivation considers a 
	plane wave laser with a temporal envelope. 
	However, one could extend this model by using a lower effective $a_0$ to account for the transverse envelope function.

	\subsection{Effects of Radiation reaction}
	%
	Radiation reaction must be present in addition to the Lorentz force to accurately resolve the motion of a particle when ultra-intense laser beams are considered. We model the radiation reaction term using the Landau-Lifshitz (LL) equation \cite{Landau1975}
	%
	\begin{equation}
	\dfrac{\mathrm{d}\mathbf{p}}{\mathrm{d}t}=-e\left(\mathbf{E}+\frac{\mathbf{p}}{\gamma m_e c}\times\mathbf{B}
	\right) +\mathbf{F}_{\mathrm{RR}},
	\end{equation}
	where the leading order term of the radiation reaction force is given by
	\begin{equation}\label{RRforce}
	\mathbf{F}_{\mathrm{RR}} =- 
	\frac{2e^{4}}{3m_{e}^{3}c^{5}}\gamma\mathbf{p}\left[\left(\mathbf{E}+\frac{\mathbf{p}}{\gamma m_e c}\times\mathbf{B}\right)^{2}-\left(\frac{\mathbf{p}}{\gamma m_e c}\cdot\mathbf{E}\right)^{2}\right]. 
	\end{equation}
	
	%
	%
	We mentioned before that the electron dynamics without radiation reaction does not depend on the value of $ f $  in 
	equation \eqref{channel_fields_eq} \cite{Huang2016}. 
	We will now show that even the radiation reaction force at the first order in $\mathcal{I}/\gamma$ does not depend on the value of $ f $.
	For simplicity, we assume again the laser to be a plane wave that is linearly 
	polarized along the $ y $-axis, and propagates along $x$. 
	We consider an electron that starts in the plane $z=0$ (in fact in the sub-plane $y>0$ without loss of generality).
	%
	
	The electromagnetic fields experienced by the electron are given by 
	equation \eqref{E_total}, where $E_\mathrm{C}$ and $B_\mathrm{C}$ are 
	positive by construction. The laser fields $E_\mathrm{L}$ and 
	$B_\mathrm{L}$ can take any sign, but they are equal to one another in 
	magnitude: $E_\mathrm{L}=B_\mathrm{L}=E_0 \sin \phi$.
	%
	%
	This field configuration cannot change $p_z$ if it was initially zero, so 
	the particle is confined in the $(x,y)$ plane. 
	We can simplify the term in square brackets of equation \eqref{RRforce} 
	considering the directions of $\mathbf{E}$ and $\mathbf{B}$ from equation \eqref{E_total}  and replace $p_x$ and $p_y$ according to the previous 
	values
	\begin{widetext}
		\begin{equation}
		\label{eq_long}
		\left(\mathbf{E}+\frac{\mathbf{p}\times\mathbf{B}}{\gamma m_e c}\right)^{2}-\left(\frac{\mathbf{p}\cdot\mathbf{E}}{\gamma m_e c}\right)^{2} \simeq 
		\left(  E_\mathrm{C}+B_\mathrm{C} + \dfrac{p_y^2}{2\gamma^2}E_\mathrm{L}  \right)^2 \simeq 
		\left( E^*_\mathrm{C} \sin \psi+\dfrac{\mathcal{I} \cos^2\psi}{\gamma} E_0 \sin \phi  \right)^2,
		\end{equation}
	\end{widetext}
	where 
	\begin{equation}
	E^*_\mathrm{C} \sin \psi=E_{\mathrm{C}}+B_\mathrm{C}\simeq\frac{m_e\omega_\mathrm{p}^2}{2e}R.
	\end{equation}
	We can draw an important conclusion from equation \eqref{eq_long}.
	In an extremely intense laser, it is likely to have $E^*_{\mathrm{C}} \ll E_\mathrm{0}$. 
	However, even though $E^*_{\mathrm{C}}$ has a significantly smaller magnitude, the channel field dominates the radiation reaction as long as $E_\mathrm{0}\ll  E^*_{\mathrm{C}}\gamma/\mathcal{I}$. 
	This inequality can be violated for a large transverse momentum component $p_y\gg p_x$. It is not excluded that some particles could be in this situation before getting accelerated in the positive $x$-direction.  
	%
	%
	
	The energy loss can be approximated by 
	\begin{equation}\label{Energy_equation}
	m_{e}c^{2}\dfrac{d \gamma}{d t}\simeq-\alpha_{\mathrm{RR}}\gamma^{2}
	\left( E^*_\mathrm{C} \sin \psi+\dfrac{\mathcal{I} \cos^2\psi}{\gamma} E_0 \sin \phi \right) ^{2},
	\end{equation}
	where $\alpha_{\mathrm{RR}}=2e^{4}/(3m_{e}^{2}c^{3})$
	is the damping constant 
	\cite{Landau1975}.
	Radiation reaction sets a limit on the maximum electron energy attainable in the interaction with an ultra-intense laser pulse.
	The limit is reached when all the energy an electron can gain during
	one betatron cycle is radiated out during the same time interval.
	As a result, the particle energy is equal before and after such a cycle.
	In other words, no further acceleration is possible. 
	For a relativistic particle that satisfies the condition $ E_\mathrm{0}\ll E^*_{\mathrm{C}} \gamma/\mathcal{I}$,
	we can estimate the maximum energy
	by equating $d\gamma/dt$ from equation \eqref{work_of_field} and 
	equation \eqref{Energy_equation} which leads to
	%
	\begin{equation}\label{gamma_max_channel0_cgs}
	\gamma_{\mathrm{RR}}\approx 
	\left(\dfrac{e^{2}a_{0}\omega_{0}}{\alpha_{\mathrm{RR}}m\omega_{\mathrm{p}}^{2}\sqrt{2\mathcal{I}}}\right)
	^{2/5}.
	\end{equation}
	Equation \eqref{gamma_max_channel0_cgs} represents the acceleration limit when the channel field is the dominant cause of radiation reaction and can be applied to quasi-resonant particles. The result can be simplified to
	\begin{equation}\label{gamma_max_channel0}
	\gamma_{\mathrm{RR}}\approx
	1.484\times10^{3}\left(\lambda_{0}[\mathrm{\mu m}] 
	\dfrac{\omega_{0}^{2}}{\omega_{\mathrm{p}}^{2}}\dfrac{a_{0}}{\sqrt{2\mathcal{I}}}\right)
	^{2/5}.
	\end{equation}
	%
	%
	%
	%
	%
	%
	\subsection{Radiation reaction-induced reduction  of $\mathcal{I}$ }
	
	When electrons lose energy due to radiation emission, the integral of motion $\mathcal{I}$ defined
	in equation \eqref{IOM} is not conserved anymore. 
	We can make a quantitative estimate of how $\mathcal{I}$ changes during one resonant betatron cycle. 
    Radiation losses change the instantaneous energy and momentum of the particles, which also reduces the integral of motion $\mathcal{I}$. 
	For relativistic particles, we can assume that most of the radiation is emitted in the direction of motion (this emission is contained within a cone that has an opening angle $\theta \sim 1/\gamma$). 
	According to this assumption, the momentum and energy are reduced 
	proportionally: $d \gamma/\gamma\simeq d p/p\simeq 
	d p_x/p_x$.
	The integral of motion therefore decreases through the reduction of 
	$\gamma-p_x/(m_ec)$, in the following way:
	\begin{equation}\label{dI}
	d \mathcal{I}=d \gamma-\dfrac{d  p_{x}}{m_{e}c} 
	\simeq d \gamma\left(1-\dfrac{p_{x}}{\gamma m_{e}c} \right), 
	\end{equation}
	where we used the proportionality mentioned above to express 
	$d p_x$ as a function of 
	$d \gamma$.
	%
    Using equation \eqref{dI} we can derive that the integral of motion is reducing at a rate
	\begin{equation}
	\dfrac{d \mathcal{I}}{d t}=\dfrac{d \gamma}{d t}~\dfrac{\mathcal{I}\cos^2\psi}{\gamma}, 
	\end{equation}
	where $d \gamma/d t$ is 
	given by equation \eqref{Energy_equation}.

	We can estimate how much $\mathcal{I}$ decreases during one resonant betatron cycle. 
	We first perform a change of variables $ 
	d t= \gamma^{*}/(\mathcal{I}\cos^{2}\psi) ~  
	d \psi$, and use equation \eqref{gamma_res} to 
	remove direct $\gamma$ dependence. 
	If we assume a limit where the channel field dominates, using $\int_0^{2\pi} \cos^8\psi \sin^2\psi ~\mathrm{d\psi}=7\pi/128$ we get $ \mathcal{I} $ decreases to
	
	\begin{equation}\label{IOM_reduced}
	\mathcal{I}=\mathcal{I}_{0}\left[ 1+3.2\times 10^{-8}\dfrac{\omega_{0}^{2}}{\omega_{\mathrm{p}}^{2}} \dfrac{\mathcal{I}_{0}^{4}}{\lambda_{0}[\mathrm{\mu 
			m}]} 
	\right]  ^{-1/4}.
	\end{equation}
	
\begin{figure}
		\centering
		\includegraphics[width=0.8\linewidth]{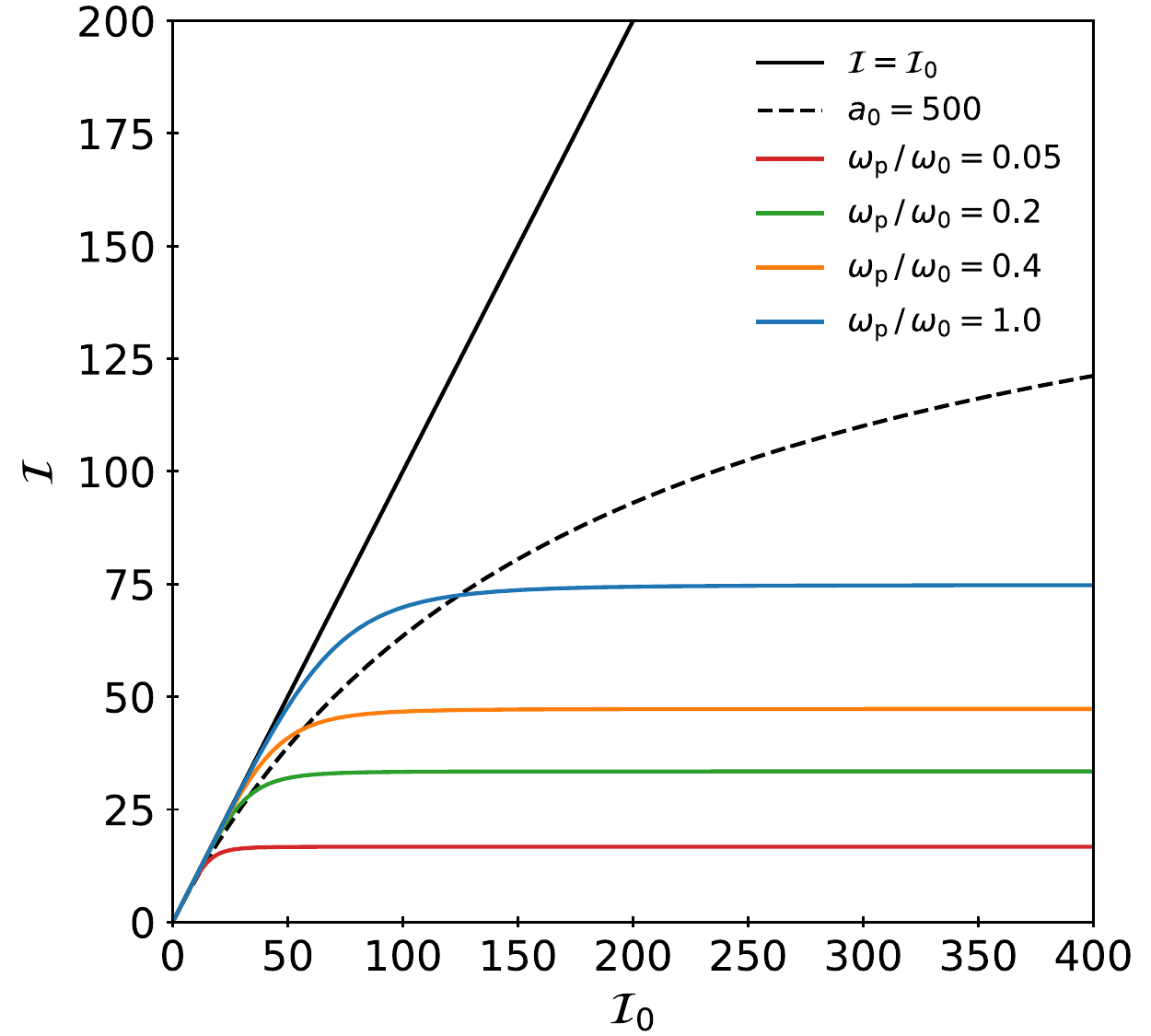}
		\caption{Reduction of the invariant $\mathcal{I}$ during one resonant betatron oscillation cycle. Dashed black line represents the prediction of equation \eqref{IOM_reduced_laser} for a laser with $a_0=500$, while the coloured lines represent the predictions of equation \eqref{IOM_reduced} for different values of $\omega_{\mathrm{p}}/\omega_0$. For reference, $\mathcal{I}=\mathcal{I}_0$ is given as a solid black line.  }
		\label{integral}
	\end{figure}

	In the other limit, where the laser field dominates, we have $\int_0^{2\pi} \cos^6\psi  ~\mathrm{d\psi}=5\pi/8$, which gives
	
	\begin{equation}\label{IOM_reduced_laser}
	\mathcal{I}=\mathcal{I}_{0}\left[ 1+2.3\times 10^{-8}~a_0^2 ~\dfrac{ \mathcal{I}_{0}}{\lambda_{0}[\mathrm{\mu 
			m}]}
	\right]  ^{-1}.
	\end{equation}
	%
	%
	%
    Both equations  ~\eqref{IOM_reduced} and  ~\eqref{IOM_reduced_laser} predict an asymptotic value that particles with a large
 	initial value of $\mathcal{I}_0$ would  tend to after a resonant cycle. 
	If the predicted asymptotic $\mathcal{I}$ is much smaller for one of the limits (RR due to the
	laser or due to the channel fields), it means that this limit represents the dominant contribution to the radiation reaction. 
	As it turns out, the $\mathcal{I}$ limit predicted by equation \eqref{IOM_reduced} is frequently lower than the one predicted by 
	equation \eqref{IOM_reduced_laser}. 
	An example for $a_0=500$ is given in figure \ref{integral}. 
   The figure illustrates that even one resonant betatron cycle can be enough to reduce the integral of motion very close to the value where the 
   radiation reaction is not very strong anymore and cannot make a significant change during the subsequent cycles. 
	Inserting the asymptotic value of $\mathcal{I}$ into equation~\eqref{gamma_max_channel0} 
	will give 
	us the maximum electron energy allowed due to the radiation reaction.
	
	For predictions of the maximum energy in the system, one should compare equations  
	\eqref{gamma_final_max},  ~\eqref{gamma_l} and ~\eqref{gamma_max_channel0}, using the lowest 
	energy predicted among the three as a final result. One should consider the $\mathcal{I}_0$ for equation ~\eqref{gamma_l}, because
	it applies to acceleration from the beginning of the interaction, while equations ~\eqref{gamma_final_max} and \eqref{gamma_max_channel0} 
	should be considered with the value of $\mathcal{I}$ reduced by radiation reaction. 

	\section{II. Test particle simulations, comparisons with the analytical model}
	
	To validate our model and probe parameters optimal for particle acceleration, we now use a test-particle 
	approach in a simplified electromagnetic configuration, similarly as in section I. We consider a wide 
	range of initial conditions varying the particle distance from the channel axis, background plasma density and the laser intensity. 
	This data can illustrate how limits presented in the previous section affect  particle acceleration.

	The laser is represented as a plane wave with a temporal envelope \cite{Huang2016,Huang2017Nonlinear},
	which propagates along the positive $ x $ direction at the speed of light and is polarised along the $y$-axis. 
	We consider  $ \lambda_{0}=1~\mathrm{\mu m} $ and a Gaussian 
	temporal profile with duration  $\tau=150~\mathrm{fs} $
	defined as full-width-at-half-maximum (FWHM)
	in the electromagnetic field amplitude. 
	Due to the slowly-varying laser envelope function, the initial 
	carrier-envelope phase does not impact the laser-electron interaction.
	%
	The test particle equations of motion 
	are integrated using the 4th order Runge--Kutta method with a time step $ \Delta t = 
	\lambda_{0}/(1000c) $ satisfying 
	the condition for accurate calculation of electron dynamics in a strong
	laser field \cite{Arefiev2016Criterion}.
	The numerical factor $ f $ introduced in equation~\eqref{channel_fields_eq} is here set to 0.5, such that the electric and magnetic field of the channel have an equal magnitude.
	The plasma channel is assumed to be axially symmetric, with a typical field structure 
	given by equation  \eqref{channel_fields_eq}.
	In our calculations, the length of the plasma channel is $ 528~\mathrm{\mu m} $.
	In the following analysis, the distances are often expressed in units normalized to the laser frequency where $ 
	c/\omega_{0} \approx 0.16~\mathrm{\mu m}$.
	The electrons always start at rest at an initial distance $R_0$ from the channel axis. 
	%
	%
	%
	%
	This  corresponds to a case when an intense laser propagates through
	a pre-formed plasma channel and  accelerates electrons originating from the channel 
	walls.
   We vary $ R_{0} $ from $0$ to $ 100\,c/\omega_{0} $ for different values of $ 
	\omega_{\mathrm{p}}/\omega_{0} $  ranging from 0.01 to 1. 
   For an optical laser, this corresponds to a plasma density range between $10^{17}$ and  $10^{21}~\mathrm{parts./cm^3}$, and 
   distance from the channel axis up to $\sim15~\mu \mathrm{m}$.
   Radius and $\omega_{\mathrm{p}}/\omega_0$ are sampled with 100 values each, which makes for $10^4$ test cases just by 
   varying these two parameters. 
   We also consider several values of $a_0$, with examples of strong and weak radiation reaction.

	\subsection{Without radiation reaction}
    
\begin{figure}
		\centering
		\includegraphics[width=1\linewidth]{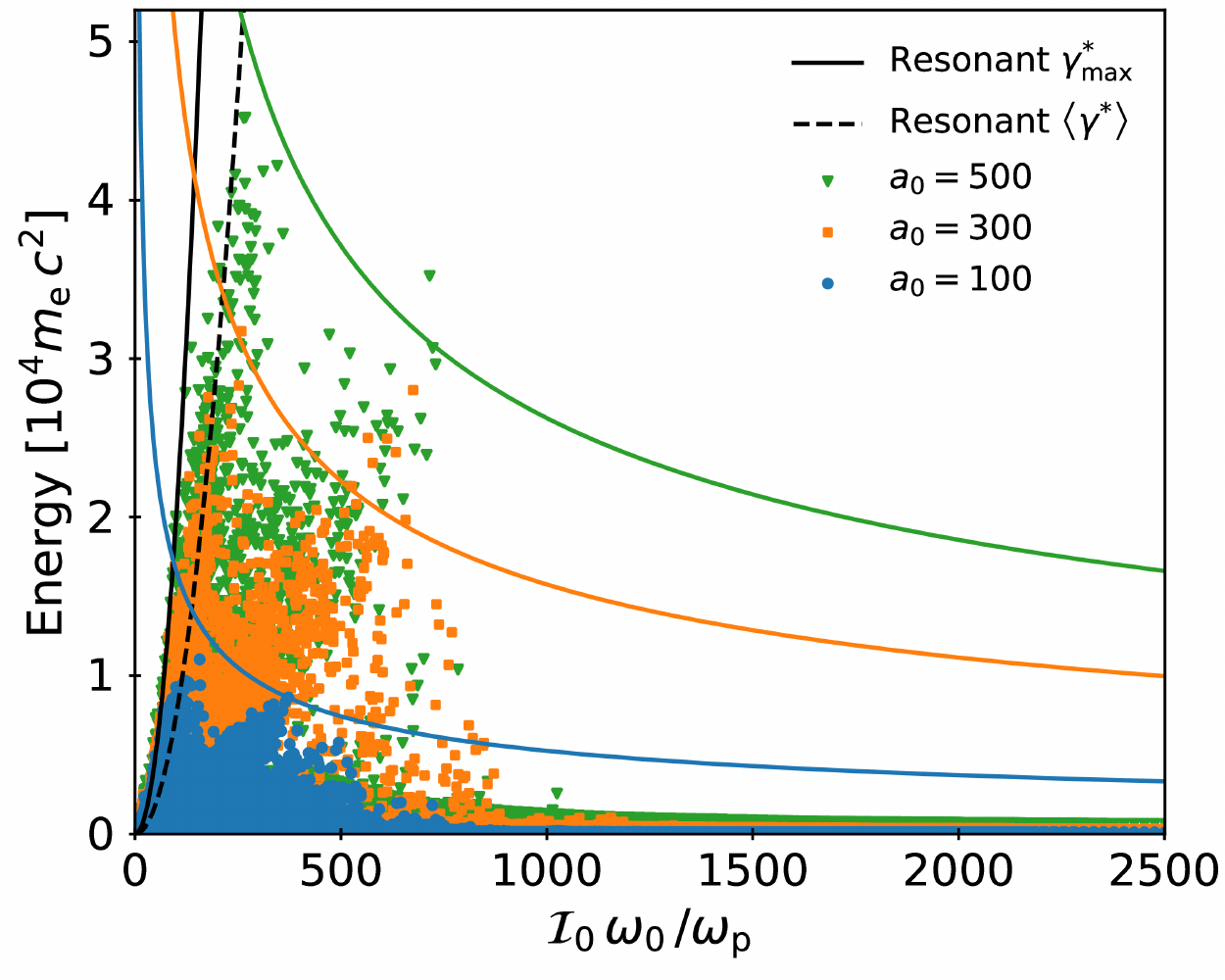}
		\caption{Data points show the maximum value of electron energy as 
			a function 
			of the integral of motion
			$ \mathcal{I}_{0} $ and the plasma frequency $ \omega_{\mathrm{p}} $ 
			for different $ a_{0} 
			$. In these runs, radiation reaction is neglected.
			Corresponding color-coded solid lines show the maximum electron 
			energy 
			according to 
			equation~\eqref{gamma_l} at the moment when the electron 
			reaches the 
			end of 
			the channel.
			The black solid (dashed) line represents the analytical 
			estimate of the 
			maximum (cycle-averaged) electron energy given by the resonance 
			condition~\eqref{gamma_final_max}.
		}
		\label{fig:03}
	\end{figure}

\begin{figure*}[!htb]
		\centering
		\includegraphics[width=0.6\linewidth]{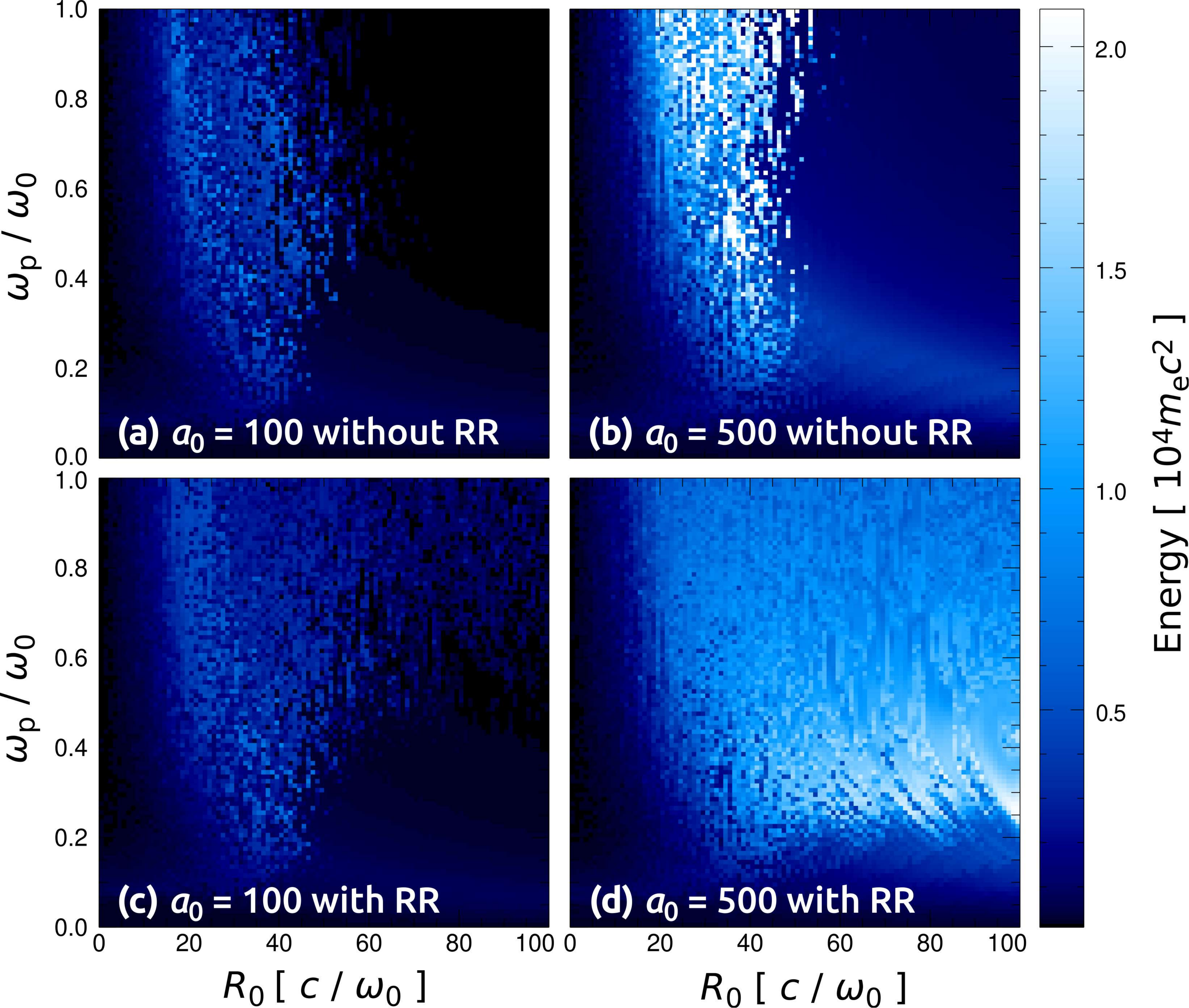}
		\caption{Maximum electron energy as 
			a function 
			of the normalized plasma frequency $ \omega_{\mathrm{p}}/\omega_{0} 
			$, the 
			initial 
			radial 
			distance $ R_{0} $ from the 
			channel axis and $ a_{0} 
			$. In cases (a) and (b), radiation reaction is neglected, while in 
			cases (c) and (d) it is taken into account.}
		\label{fig:04}
	\end{figure*}

    Figure~\ref{fig:03} summarizes the results obtained from  
	test-particle 
	simulations without radiation reaction for three different values of $a_0$. 
	Each data point represents the maximum achieved value of relativistic Lorentz factor $ \gamma $ of an 
	electron as a function of the initial value for $\mathcal{I}_0 ~\omega_{\mathrm{0}}/\omega_{\mathrm{p}}$.
	%
	%
  Some particles achieve energies over 10 GeV, and many of them are resonant
	according to equation \eqref{gamma_res}. 
	This is verified in figure \ref{fig:03}  where $ \gamma^{*}_{\mathrm{max}}  $ is shown as a black solid line, while cycle-averaged value  $ \left\langle 
	\gamma^{*} \right\rangle $ is represented with a dashed line. 
	Particles are resonant at different energies for different values of $\mathcal{I}_0 ~\omega_{\mathrm{0}}/\omega_{\mathrm{p}}$, and, as shown in section I, the resonance condition does not depend on the laser intensity. 
	The asymptotic energy, therefore, has to be estimated by considering how much work towards particle acceleration the laser can 
	invest during the available acceleration distance. 
	According to equation ~\eqref{gamma_l}, this limit is a function of $a_0$, $\mathcal{I}_0$ and $\omega_{\mathrm{p}}$. 
	The color-coded solid lines show the maximum possible electron energy for quasi-resonant particles given by 
	equation~\eqref{gamma_l} corresponding to the acceleration distance of 528 $\mu \mathrm{m}$.
   For a specific set of parameters $a_0$ and $\omega_{\mathrm{p}}$, the most favourable initial $R_0$ for acceleration is the 
	one that allows to achieve the betatron resonance at a highest  electron energy within the available acceleration distance. 
	This optimal $R_\mathrm{opt}$ corresponds to the intersection between the dashed and the coloured solid curves in figure \ref{fig:03}, and is given by equation ~\eqref{optimal_r}.
	The acceleration is in principle possible even far from betatron resonance, with a slightly lower maximum allowed energy. 
	However, as we have mentioned before, the probability for constructive work performed by the laser towards particle acceleration is low for high values of $\mathcal{I}_0$, where $R_0\gg R_\mathrm{opt}$. 
    The existence of an optimal initial radial position is better illustrated in figure~\ref{fig:04}, where panels (a) and (b) display the test electron maximum energy as a function of $R_0$ and $\omega_{\mathrm{p}}/\omega_{0}$ for the same dataset summarized in  figure \ref{fig:03}.  
	%
	%
	We can also see sudden jumps in the maximum achieved  energy between points with similar parameters, which 
	suggest a resonant nature of the acceleration mechanism, as highlighted in the previous works 
	\cite{Pukhov1999,Tsakiris2000}.
	%
	%
	%
	%
	%
	%
	%
	%

	\subsection{Effects of radiation reaction}
	
	When considering extreme laser intensities, we expect radiation reaction 
	to be important for the particle dynamics, and consequently for acceleration.
	This section addresses the radiation reaction effects. 
     We present  test particle simulations for the same parameters as in the previous section, but now including RR in the equations of motion. 
	In the code, RR is implemented as a continuous 
	process 
	of losing electron energy described by the Landau-Lifshitz equation \cite{Landau1975,Vranic2016}.  
	We do not include the electron Gaunt factor correction in these simulations - we verify later in this section that this does not change our findings.
	The results of test-particle simulations for $ a_{0}=100 $ and $ a_{0}=500 $ are shown in figure~\ref{fig:04} (c) and (d).
	%
	%
	%
	The differences between cases with and without RR for $a_0=100$ are small, both in the 
	absolute values of the energy obtained as well as in the parameters favourable to achieve betatron resonance. 
	This suggests that RR is not strong for $a_0=100$. 
    In contrast, the maximum energy achieved with RR is about a factor of two lower than without RR for $a_0=500$. 
	But the most striking difference is in the parameter range where the maximum energy is achieved. 
	It appears that all particles with $R_0>R_\mathrm{opt}$ can get accelerated. 
	Since radiation reaction affects the electron dynamics and thus the evolution of 
	the phase-matching process between the particle oscillations in the channel and the laser field, 
	an efficient electron acceleration can be achieved for a wider 
	range 
	of initial conditions. 
	%
	%
	%
	This is accomplished through the reduction of the integral of motion, which allows particles with 
	an initially large $\mathcal{I}_0$ to eventually converge to a lower quasi-resonant value of $\mathcal{I}$ that 
	enables energy retention. 
	
	
	\begin{figure*}
		\centering
		\includegraphics[width=1\linewidth]{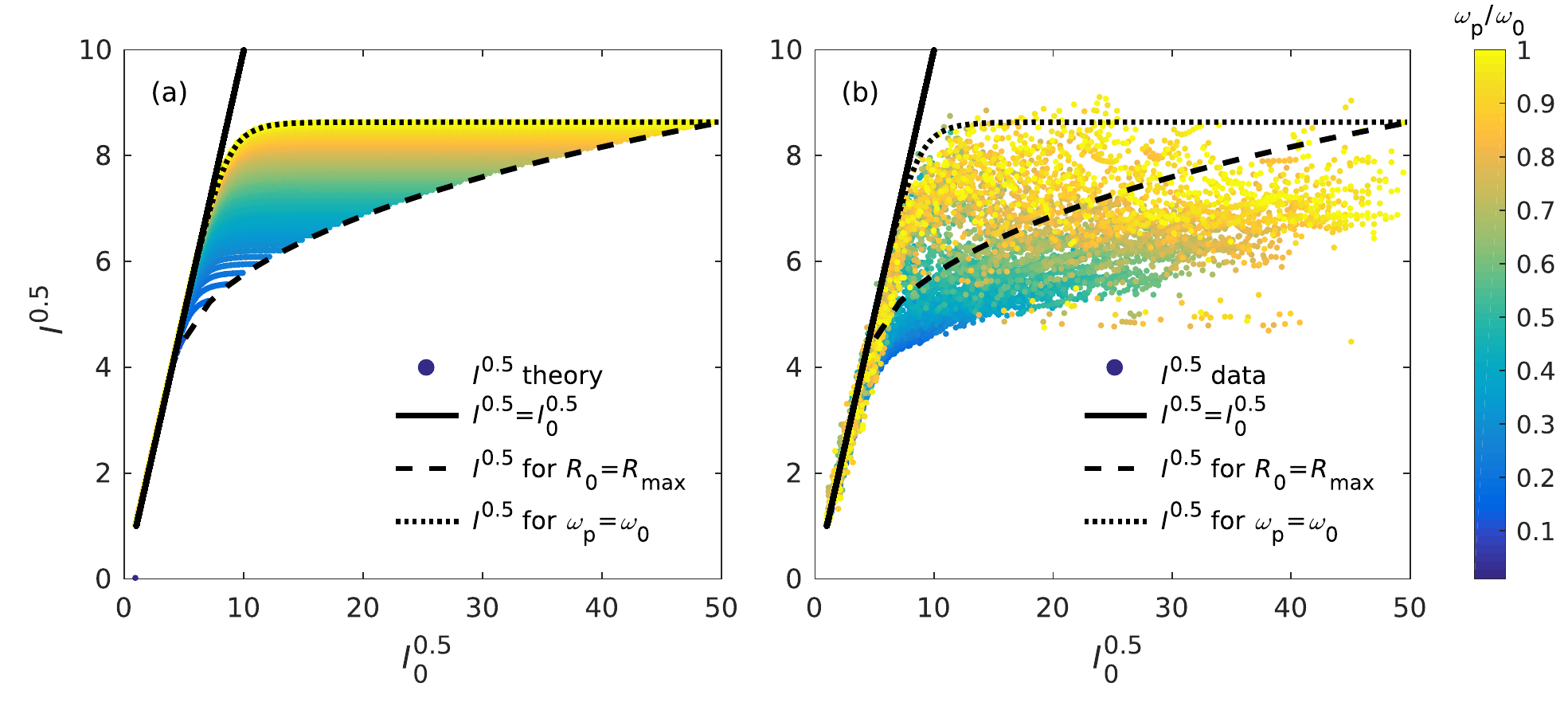}
		\caption{(a) Theoretical estimate of the reduction of the  integral of 
			motion $\mathcal{I}$  after one 
			resonant cycle given by equation~\eqref{IOM_reduced}. (b) Comparison 
			with 
			data points from test-particle simulations  at the moment when the electron reaches the maximum energy. The considered laser intensity for both panels corresponds to $ a_{0}=500$.}
		\label{fig:07}
	\end{figure*} 
	
	\begin{figure*}
		\centering
		\includegraphics[width=1\linewidth]{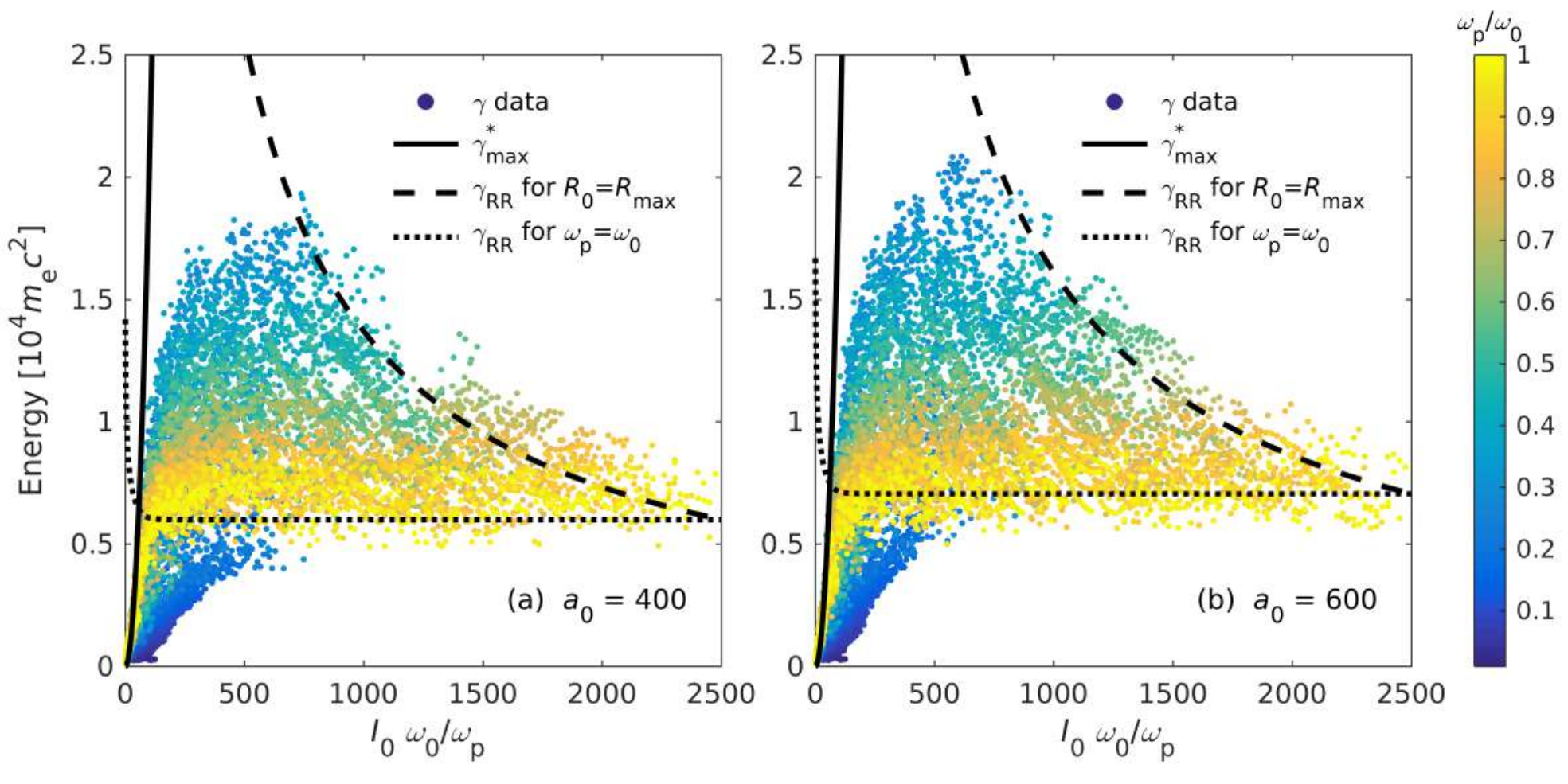}
		\caption{Summary of parameter studies for (a) $ 
			a_{0}=400 $ and (b) $ 
			a_{0}=600 $ with $ 10^{4} $ test
			cases varying the plasma frequency $ \omega_{\mathrm{p}}/\omega_{0} $ and
			initial radial electron position $ R_{0} $ when radiation reaction is taken 
			into account. The solid line corresponds to $ 
			\gamma^{*}_{\mathrm{max}} $, 	
			equation~\eqref{gamma_final_max}, dashed and dotted 
			line
			to $ \gamma_{\mathrm{RR}} $, 
			equation~\eqref{gamma_max_channel0}, for $ 
			R_{0}=R_{\mathrm{max}}=100\,c/\omega_{0} $ and for 
			$ \omega_{\mathrm{p}}=\omega_{0} $, 
			respectively. 
		}
		\label{fig:06}
	\end{figure*}
	
		\begin{figure*}
		\centering
		\includegraphics[width=1\linewidth]{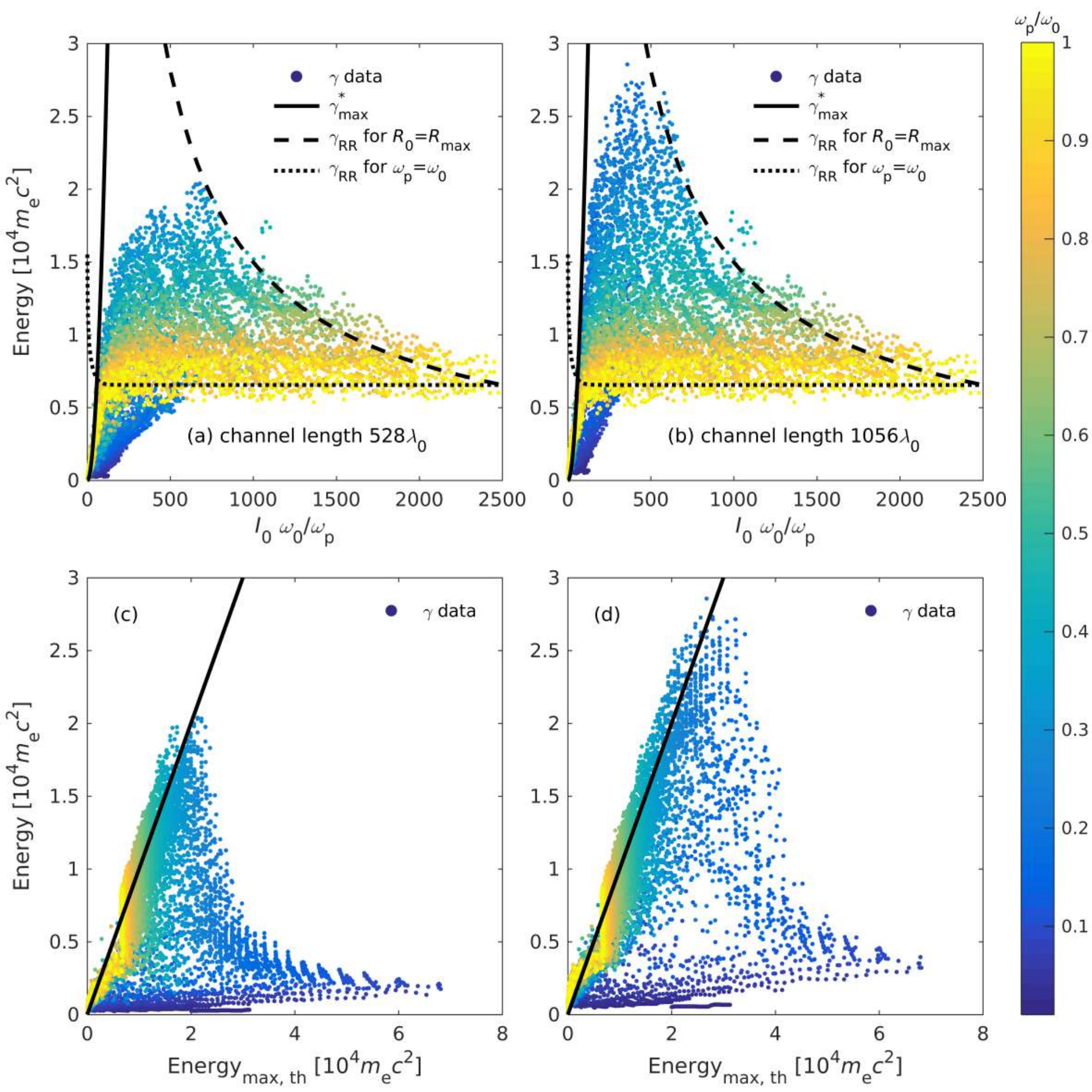}
		\caption{
			Maximum electron 
			energy for $a_0=500$
			in case of (a), (c) $ 528~\mathrm{\mu m} $ and (b),(c) $ 
			1056~\mathrm{\mu m} $
			long 
			channel.
			(a), (b)  Solid line corresponds to $ \gamma^{*}_{\mathrm{max}} $, 	
			equation~\eqref{gamma_final_max}, dashed and dotted line
			to $ \gamma_{\mathrm{RR}} $, 
			equation~\eqref{gamma_max_channel0}, for $ 
			R_{0}=R_{\mathrm{max}}=100\,c/\omega_{0} $ and for 
			$\omega_{\mathrm{p}}=\omega_{0} $, respectively. (c), (d) Achieved energy vs. maximum allowed energy for same datasets as in (a), (b). Solid lines represent cases where the theoretical prediction is equal to the energy attained in the simulations.}
		\label{fig:08}
	\end{figure*}
	
	We have compared the 
	theoretical 
	estimate for reduction of $\mathcal{I}$ during one resonant cycle with the data obtained from test-particle 
	simulations.
	The predictions of equation \eqref{IOM_reduced} against the data for $ a_{0}=500 
	$ are shown in 
	figure~\ref{fig:07}.
	The values of $\mathcal{I}$ for our parameter range are restricted by three boundaries, as shown in figure~\ref{fig:07} (a).
	The solid line represents a situation of weak radiation reaction where the integral of motion does not change.
	%
	%
	The dotted line corresponds to the highest plasma density we considered $\left( \omega_{\mathrm{p}}/\omega_{0}=1 
	\right)$, while the dashed line is for electrons starting at the furthest initial position from the channel axes.  
	Figure \ref{fig:07} (b) shows the integral of motion calculated from the data 
	 at the moment when the electron reaches its maximum energy.
	Equation \eqref{IOM_reduced} is a good predictor for the order of the final value of $\mathcal{I}$, even if we consider
	only one resonant betatron cycle. 
	Some particles will have enough time to perform multiple betatron oscillations, and achieve the 
	asymptotic value of $\mathcal{I}$ where the radiation reaction becomes negligible. 
	However, since this value is of the same order, for simplicity we will continue considering a single betatron cycle for computing the final $\mathcal{I}$. 
    After inserting this corrected value of $\mathcal{I}$ into equations ~\eqref{gamma_final_max} and \eqref{gamma_max_channel0}, the equation that predicts
    a lower energy is a good estimate of the asymptotic energy of the electron, provided that the acceleration distance was long enough to achieve this energy according to equation \eqref{gamma_l}.
	%

	%
	%
	%
	%
%
	%
 We can verify this by presenting the test particle simulations with radiation reaction  
 in a similar fashion as in figure \ref{fig:03}. Figure \ref{fig:06} shows the maximum
 particle energy as a function of $\mathcal{I}_0 ~\omega_{0}/\omega_{\mathrm{p}} $ 
 for two different laser intensities, 
  $ a_{0}=400 $ and $a_{0}=600 $. 
  %
	%
	If we assume that the electron is initially at the maximum distance $ 
	R_{\mathrm{max}}=100\,c/\omega_{0} $ from the 
	channel axis, then equation \eqref{gamma_max_channel0} provides the limit on 
	the attainable electron energy which is illustrated by the 
	dashed line in figure \ref{fig:06}.
	The dotted line represents the expected value for $\omega_{\mathrm{p}}=\omega_0$ according to equation \eqref{gamma_max_channel0}.
For $\omega_{\mathrm{p}}/\omega_{0}<0.3$, there is  a window of parameters that allows energies
above 10 GeV. 
However, we do not see them in our results, because of the finite interaction time. 
As has been stated in the previous subsection, longer acceleration distance 
allows  to accumulate more energy.
This is illustrated in figure~\ref{fig:08}, where all parameters are kept equal, except the 
acceleration distance.
The upper panels show the same plots as figure \ref{fig:07} for $a_0=500$ and $l_\mathrm{acc}=528~\mu\mathrm{m}$ or $l_\mathrm{acc}=1056~\mu\mathrm{m}$. 
The lower panels show the obtained energy vs. expected energy according to the minimum predicted by equations
  \eqref{gamma_final_max} and \eqref{gamma_max_channel0}.
%
There is a clear convergence of the data towards the maximum allowed energy we predict, but there are also outliers, especially for $\omega_{\mathrm{p}}/\omega_0<0.3$.
These particles do not attain their allowed maximum, either because there is not enough time, or because their initial integral of motion is already so low that 
the radiation reaction does not significantly affect their motion and they can easily remain far from the betatron resonance. 
A fraction of these particles eventually does converge towards the predicted values in the example with a longer propagation distance. 

\subsection{Approximations: limits and validity}

	We have introduced several approximations in our analytical model, that we hereby justify.

\begin{figure*}[!htb]
			\centering
			\includegraphics[width=0.6\linewidth]{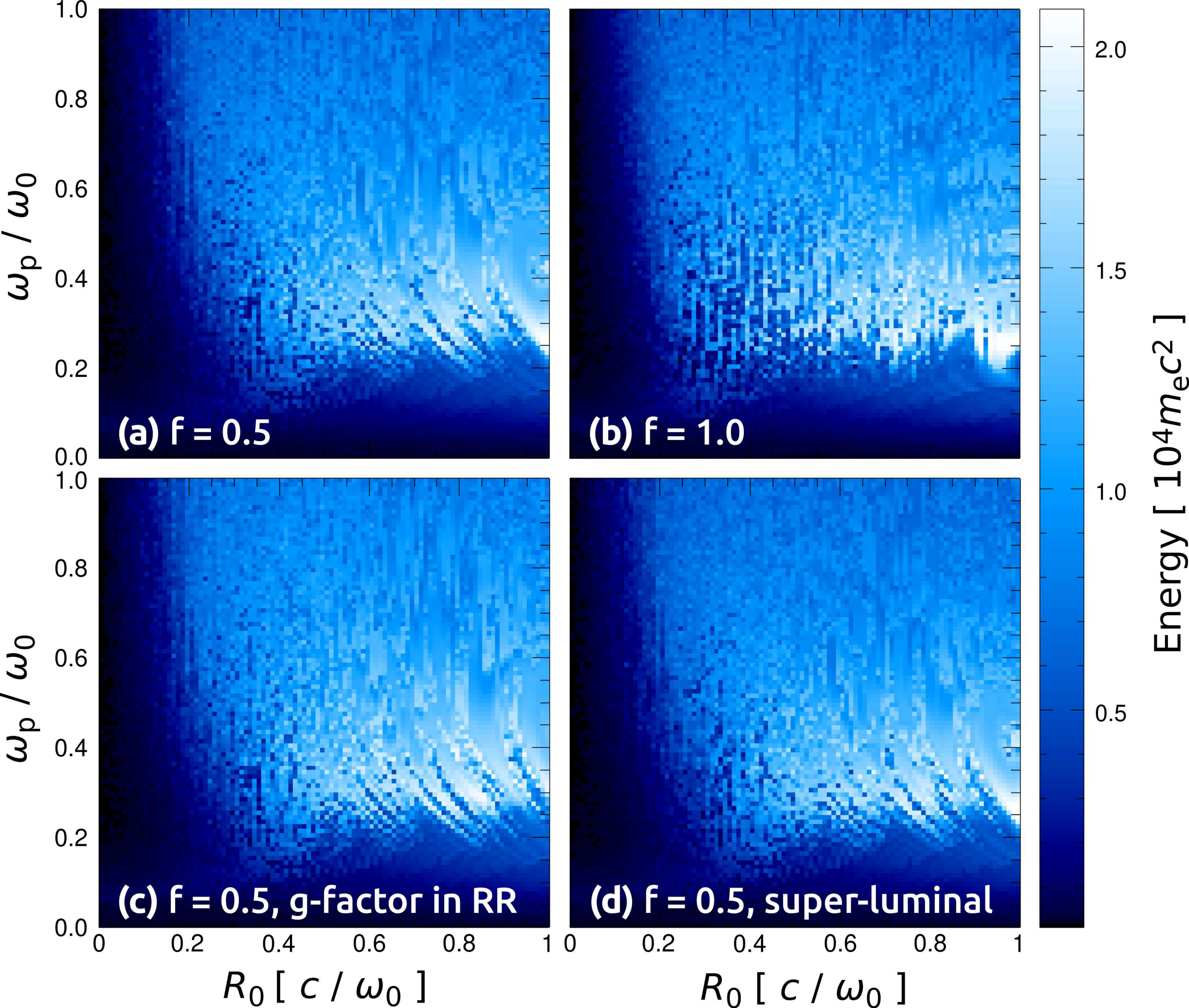}
			\caption{Maximum electron energy as 
				a function 
				of the normalized plasma frequency $ 
				\omega_{\mathrm{p}}/\omega_{0} 
				$ and the 
				initial 
				radial 
				distance $ R_{0} $ from the 
				channel axis for $ a_{0} = 500
				$ with radiation reaction.
				(a) Parameter $ f $ in equation \eqref{channel_fields_eq} is 
				set to 0.5, the velocity of the wave is luminal and the 
				electron Gaunt factor is not considered. (b) Same as panel 
				(a), except that the value of parameter $ f $ is now set to 
				1.0. 
				(c) Gaunt 
				factor is included in the particle equations of motion, $ f=0.5 $, the wave is luminal. (d) The wave is super-luminal, $ f=0.5 $, Gaunt factor is neglected.}
			\label{fig:sps}
\end{figure*}

    The first question that might arise is that of the choice 
    of numerical factor $f$ in equation \eqref{channel_fields_eq} that 
    represents the magnitude of electric versus magnetic field of the channel. 
    The analytical model does not predict changes in the expected energies for 
    any value of $ f $ between 0 and 1, and one can verify numerically that 
    this is indeed the case. Examples for $f=0.5$ and $f=1.0$ are shown on a 
    parameter study for $a_0=500$ in figure \ref{fig:sps} (a) and (b). There 
    are differences for individual test particles (represented by individual 
    pixels), but the expected maximum energy for a given region of parameter 
    space remains unchanged.  However, if the 
    electric field turned out to be very low (i.e. $f\rightarrow0$), particles would 
    need an initial transverse momentum to be efficiently accelerated, because 
    there would be no electrostatic potential in the $y$-direction.

	For simplicity, we have neglected the semi-classical correction to the radiation reaction, modelled by the Gaunt factor $ g(\chi_{e}) $ that reduces the amount of emitted power as a function of parameter $ \chi_{e} $. 
	The parameter $ \chi_{e} $ characterizes the interaction of an electron 
	with a
	strong electromagnetic field \cite{Nikishov1964,Zeldovich1975}, and it 
	is 
	approximately equal to the ratio of the field 
	as seen by the electron in its rest frame to the Schwinger limit 
	defined by 
	$E_\mathrm{S}=m_e^2c^3/(\hbar e)$. 
	Our analysis considered only the leading order term of the radiation 
	reaction, while the correction function is the next order in $ \chi_{e} $ \cite{Ritus1985}. The final electron energies within the channel depend on where the radiation reaction shuts down, and therefore, the most important section of the trajectory should be in the regime where $\chi_e\ll 1$. In that case, the g-factor is a very small correction.
	We have verified this by introducing the g-factor and showing the expected 
	electron energies on a parameter study for $a_0=500$ in figure 
	\ref{fig:sps} (c). The differences from results in panel (a) are very 
	small, which confirms this approximation was adequate.

    The phase 
    velocity of the laser propagating through plasma is considered luminal in our model. The actual phase velocity can be expressed as $v_{\mathrm{ph}}\simeq 
    c~\left[ 1+n_\mathrm{p}/(n_\mathrm{c} a_0)\right] $, where  
    $n_{\mathrm{c}}=\omega_{\mathrm{0}}^{2}m_{e}/(4\pi 
   e^2)$ represents the non-relativistic critical plasma density. This 
   evidently becomes important for dense plasmas and moderate laser 
   intensities. However, as phase difference accumulation between the particle 
   and the wave is an important factor in the acceleration process, there is a 
   possibility that even a small change in the phase velocity of the laser may 
   be relevant for particle dynamics. The luminal approximation is justified as 
   long as the phase accumulation due to the transverse motion of the particle 
   is much larger than the effect of super-luminosity. This is equivalent to 
   stating that the difference between the speed of light and particle average 
   velocity in the direction of laser propagation $c-\langle v_x \rangle$  is 
   much larger than $v_{\mathrm{ph}}-c$. We can estimate using $p_x$ from 
   section I, that $\langle v_x \rangle\simeq c\left[ 1-(1/2)(\langle p_y 
   \rangle/(\gamma m_e c))^2\right] $ which can be approximated as $\langle v_x 
   \rangle\simeq c\left[ 1-(1/2)(\mathcal{I}/\gamma)\right] $. To neglect 
   superluminosity, 
   the following condition must be met $\mathcal{I}/(2\gamma)\gg 
   (\omega_\mathrm{p}/\omega_0)^2/a_0$. As $\mathcal{I}$ reduces, and $\gamma$ 
   increases 
   over time, the left-hand side of the inequality has a lowest value in the 
   asymptotic conditions, where the particles have reached their maximum 
   allowed energy. We can then use the expression for $\langle \gamma^*\rangle$
   from equation \eqref{gamma_final_max}, as well as $\mathcal{I}\sim 50 
   \sqrt{\omega_\mathrm{p}/\omega_0}$ according to equation 
   \eqref{IOM_reduced}. The 
   condition that must be satisfied then becomes $a_0^2\gg 5\times10^3 
   \left(\omega_\mathrm{p}/\omega_0 \right)$. The condition is verified for our 
   parameters, which can also be confirmed from figure \ref{fig:sps}~(d). 
   Therefore, similarly as in reference \cite{Gong2019} we conclude superluminosity 
   is a minor effect in our conditions. According to this estimate, for optical lasers $a_0\gtrsim300$, the superluminal phase velocity would become important for targets of solid density $n_\mathrm{p}\gtrsim 100~n_\mathrm{c}$, and negligible for gas jets where $n_\mathrm{p}\sim n_\mathrm{c}$. 

\section{III. Particle-in-cell simulations}

\begin{figure}
	\centering
	\includegraphics[width=1\linewidth]{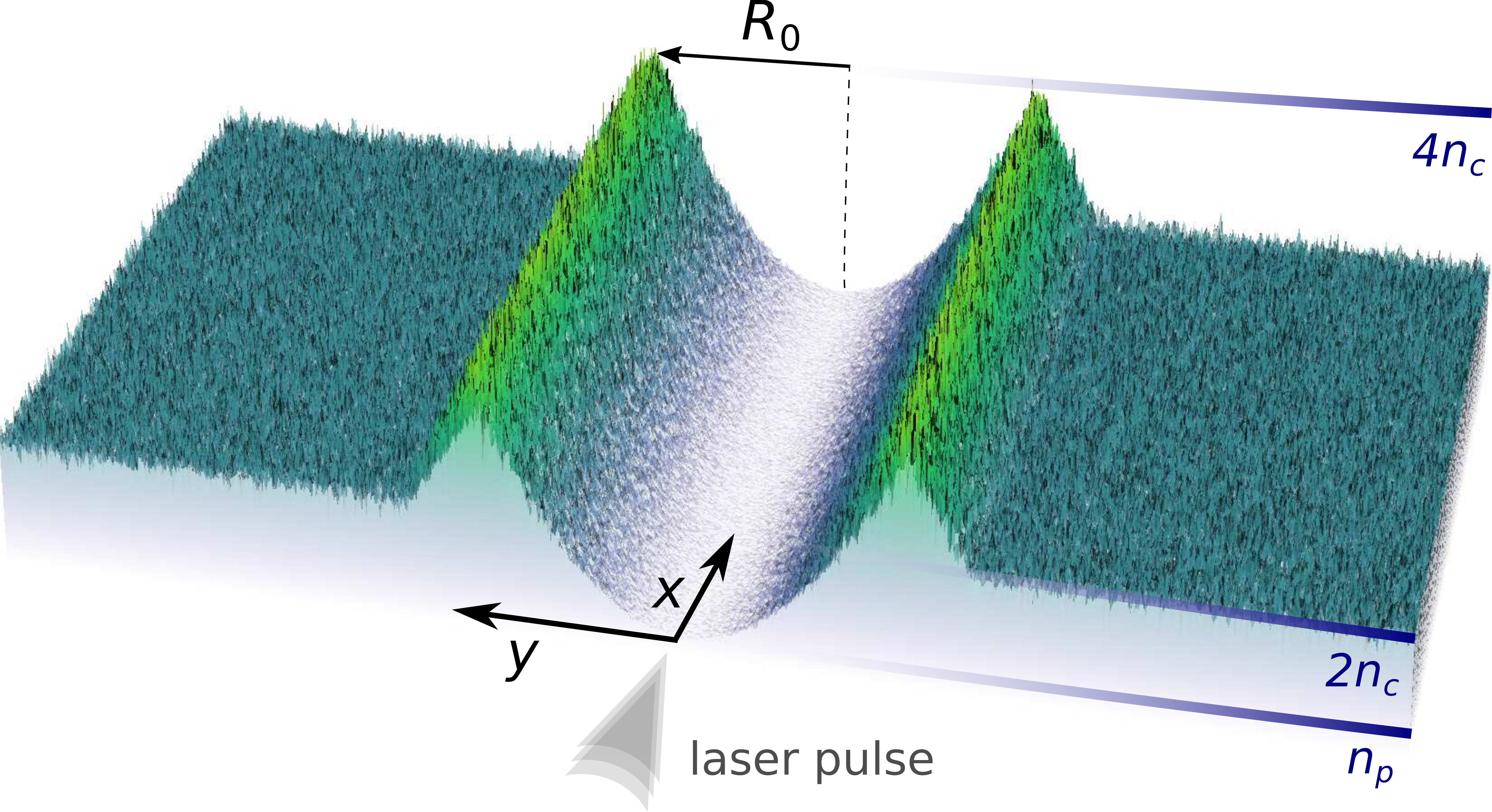}
	\caption{Transverse profile of the pre-formed plasma channel in PIC simulations.
   Plasma density at the channel axis is $n_\mathrm{p}$. 
   At the distance $ R_{0} $, the plasma density reaches its peak $ 
   4 n_{\mathrm{c}}$, while outside the channel it is set to $ 2n_\mathrm{c} $, 
   where $n_{\mathrm{c}}$ is the critical plasma density. The incoming laser pulse is propagating 
   along the positive $ x $-direction.
	}
	\label{channel_sketch}
\end{figure}

	We also compare the maximum electron energy predicted by theory against
	2D PIC simulations performed with OSIRIS \cite{Fonseca2002,Fonseca2013}.
	We consider several examples of initial conditions for $ R_{0} $ and $ \omega_{\mathrm{p}}/\omega_{0} $, and two values for the laser intensity $ a_{0}=400 $ and $ 
	a_{0}=600 $. 
	The simulation domain for PIC examples was $ 
	278~\mathrm{\mu m}\times286~\mathrm{\mu m} $ resolved with $ 17~ 500\times18~ 000 
	$ 
	cells.
   The simulations start with a pre-formed, fully ionized hydrogen plasma 
   channel illustrated in figure~\ref{channel_sketch}. 
   Such a channel can be generated by a preceding laser 
   pulse, as presented in reference~   
   \cite{Vranic2018}.
   The existence of a plasma channel is beneficial for laser guiding and can allow for longer propagation distances than if we rely on self-focusing.
   %

%
   Plasma density at the channel axis is 
   $n_\mathrm{p}$. 
   At the channel radius $ R_{0} $, the plasma density reaches its peak $ 
   4 n_{\mathrm{c}}$, while outside the channel it is set to $ 
   2n_\mathrm{c} $.
   The protons are mobile and have the same initial density profile as the electrons.
   The total length of the plasma slab is $l_\mathrm{acc}=528~\mu \mathrm{m}$, and there is a $ 5.5~\mathrm{\mu m} $  ramp at the entrance and the exit of the plasma. 
   We use the moving window, and periodic transverse boundary conditions.
  The laser duration is 150 fs, defined at FWHM in the electromagentic field amplitude.
  The transverse profile is Gaussian, with a focal spot radius $ 
  W_{0}=3.2~\mathrm{\mu m} $,  and the laser wavelength is 
  $\lambda_0=1~\mu\mathrm{m}$.
  Any effects potentially associated with super-luminal phase velocity of the laser are naturally included in the PIC simulations.
  We will show this later.
  The radiation reaction is incorporated as 
follows: for $ \chi_{e}\leq0.2 $ we use the Landau-Lifshitz equation \cite{Landau1975} with implementation details given in reference \cite{Vranic2016}. For  $ 
\chi_{e} > 0.2 $ it is modeled with a QED Monte-Carlo based algorithm \cite{Thomas_POP_2016,ThomasQED}.
The correction to the classical radiation reaction due to the electron Gaunt 
factor was not considered in the PIC simulations, as it is not expected to 
affect our conclusions according to Section II.
%

\subsection{Energy cutoff}

In the following, we compare the highest energy achieved in PIC simulations with the scalings presented in section I.
 In one PIC simulation, parameters $ a_{0} $ and $ \omega_{\mathrm 
 {p}}/\omega_{\mathrm{0}} $ are fixed, but particles can have different $ R_{0} $, and the 
 maximum available integral of motion $ \mathcal{I}_0 $ will depend on the 
 channel radius and laser spotsize.
 According to sections I and II, the particles with the maximum initial $ \mathcal{I}_0$ will have the best conditions for acceleration. We therefore use the channel radius to calculate the characteristic  $ \mathcal{I}_0 $ which gives us the estimate for the expected cutoff energy 
 in a PIC simulation.

	%
	%

	%

	%
	
	%
	%
	%

\begin{figure}
		\centering
		\includegraphics[width=1\linewidth]{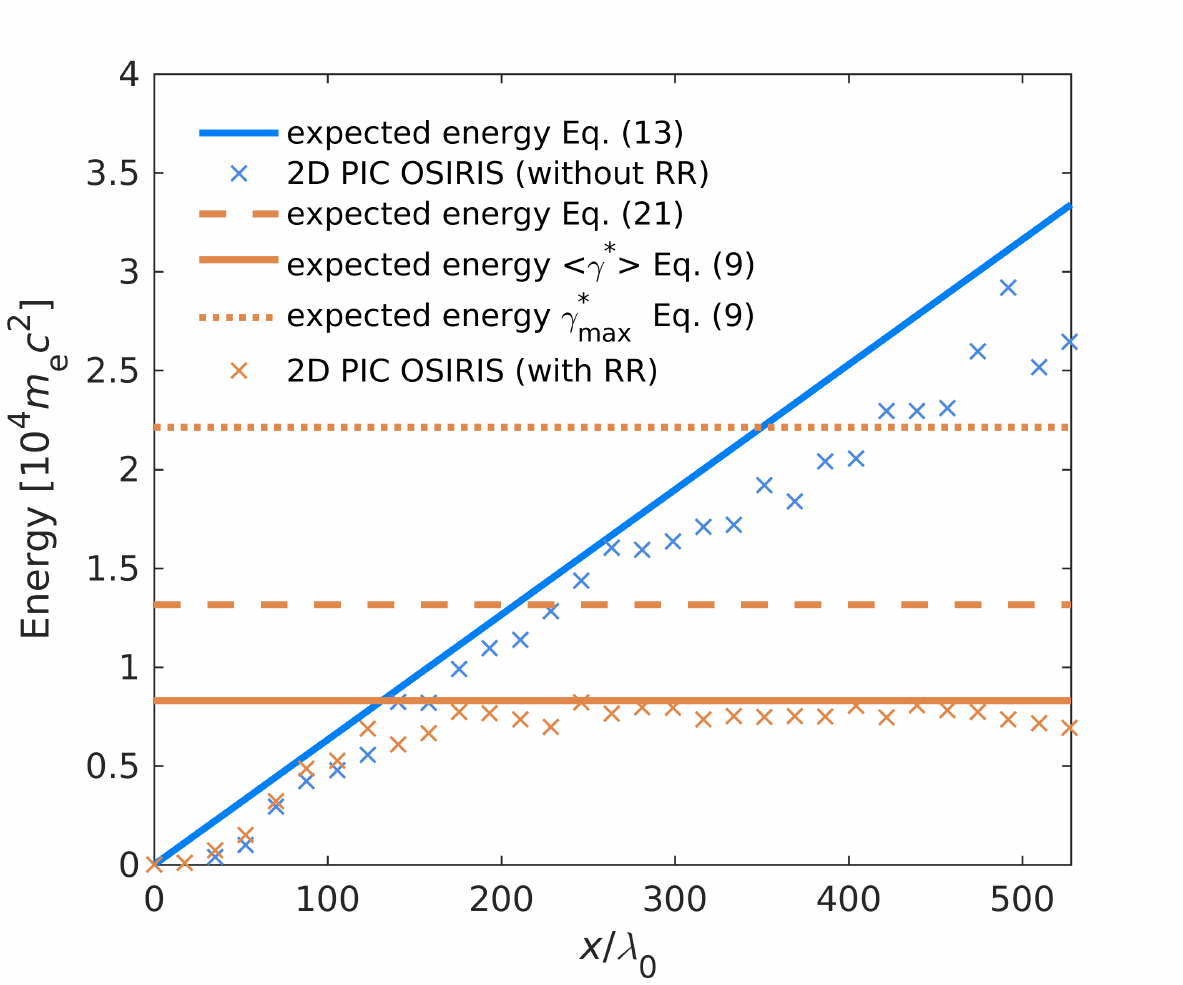}
		\caption{The maximum electron energy as a function of the propagation distance 
			for the case without (blue) and with (orange) radiation reaction for $ 
			\omega_{\mathrm{p}}= 0.5~\omega_{0}$, $R_{0}= 
			47\,c/\omega_{0} $ and $ 
			a_{0}=600 $. Solid blue line 
			represents the analytical estimate given by equation~\eqref{gamma_l} without 
			radiation reaction (considering $\mathcal{I}_0$). Orange lines are obtained using equations
			\eqref{gamma_final_max} and \eqref{gamma_max_channel0} with radiation reaction (considering $\mathcal{I}$ given by equation ~\eqref{IOM_reduced}). Crosses show the 
			corresponding 
			results from 2D PIC simulations. 
		}
		\label{fig:09}
	\end{figure}

	%
Figure~\ref{fig:09} shows how maximum energy in the simulation box grows as 
 a function of the propagation distance, both with and without radiation reaction. 
%
%
If radiation reaction is neglected, the electron energy should scale according to equation~\eqref{gamma_l}.
This trend is confirmed by the results from 2D PIC simulations.
However, radiation reaction limits the energy gain. 
%
%
Predictions of equations  \eqref{gamma_final_max} and \eqref{gamma_max_channel0} are shown together with  the data from the 2D PIC simulation with radiation reaction  that lies below all limits.
As expected, the energy cutoff is lower with radiation reaction but the number of accelerated particles is higher. For example, at $t=210~T_0$, shortly after the radiation reaction effects become visible in global quantities, the simulation with radiation reaction has 2.5 times more particles with energy above 2 GeV than the simulation without radiation reaction, even though the latter has a cutoff energy nearly twice as high.

\begin{figure*}
	\centering
	\includegraphics[width=1\linewidth]{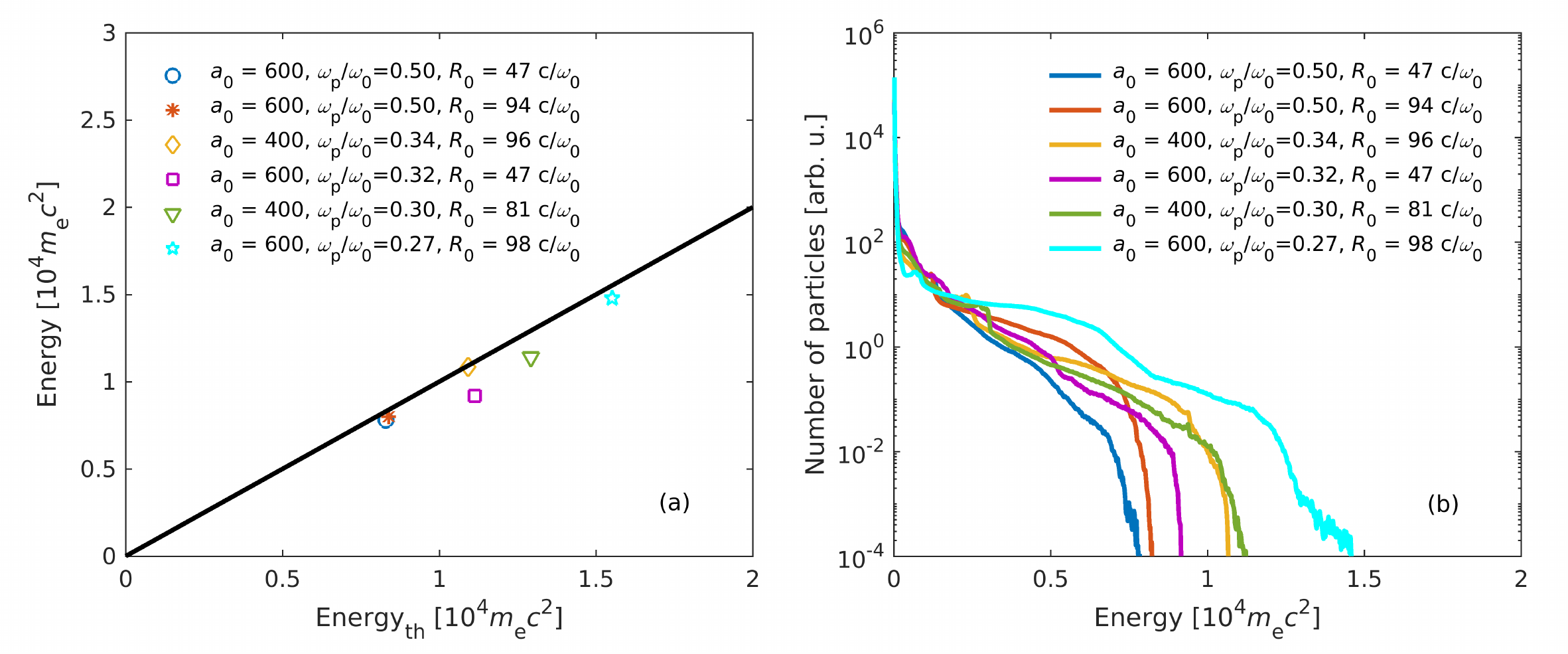}
	\caption{(a) Achieved energy in PIC simulations vs. 
	theoretical 
	expectation. 
	The theoretical prediction is
		obtained by taking a minimum predicted value using equation 
		\eqref{gamma_max_channel0},  $ \langle \gamma^{*} 
		\rangle $ from equation \eqref{gamma_final_max} and 
		equation \eqref{gamma_l}. Solid line represents cases where the 
		theoretical prediction is equal to the energy attained in the 
		simulations. (b) Electron energy distribution at the end of the channel 
		in corresponding PIC simulations.
	}
	\label{PIC_all}
\end{figure*}

Among the analytical limits, equation \eqref{gamma_final_max} predicts the lowest expected value for the considered parameters. 
According to our previous analysis, we would expect in this case to have an energy cutoff given by  $\gamma^*_\mathrm{max}$. 
Instead, the data with radiation reaction converges towards the cycle-averaged value $\langle \gamma^*\rangle$, which is lower. 
This is not surprising, because our previous analysis considered ideal conditions: the laser transverse intensity distribution was neglected, as well as the energy depletion during  propagation. 
In a realistic setup, particles do not feel the maximum laser intensity everywhere in the channel, and they do not necessarily originate exactly from the edges of the channel. 
The channel fields are a linear function of $R$ just in a region close to the channel axis,  and we have no guarantee that particles initially further than a certain threshold radius
would be accelerated because of the finite transverse laser profile. 
The asymptotic $\mathcal{I}$ can also be lower than the prediction of equation \eqref{IOM_reduced} applied to the $\mathcal{I}_0$ that corresponds to the channel radius. 
All these effects cannot be formally taken into account for analytical predictions. 
But if we probe several different combinations of parameters, taking  $\langle \gamma^*\rangle$ instead of $\gamma^*_\mathrm{max}$ provides a very good agreement with the energy cutoff measured from 2D PIC simulations. 
This is shown in figure \ref{PIC_all} (a).
The energy distribution of the electrons 
is presented in figure \ref{PIC_all} (b).
The first two examples illustrate how channel radius changes the electron distribution function. They have the same laser intensity $a_0=600$ and the same background plasma density for which $\omega_\mathrm{p}/\omega_0=0.5$, but the channel radius changes by a factor of two. The energy cutoff does not change significantly, which suggests that the narrower channel is wide enough for electrons to achieve the maximum allowed energy. However, the number of particles that approach this value is significantly higher in the case with the wider channel, which is reflected on the slope of the spectrum. Other examples show that the cutoff can be increased by reducing the plasma density. The  optimal initial distance from the axis depends on the plasma density according to equation \eqref{optimal_r}, so reducing the density to $\omega_\mathrm{p}/\omega_0=0.32$ does not increase much the energy cutoff if the channel radius remains small $R_0=47~c/\omega_0$. However, if one combines the wider channel with the lower plasma density, the electron energy cutoff evidently increases. By applying this strategy, it is possible  to obtain more energetic electrons even using lower laser intensities (e.g. $a_0=400$ in figure 13).

\subsection{The electromagnetic field structure}

We proceed to a more detailed analysis of PIC results, to illustrate the connection with the analytical model presented in Section I.
In figure~\ref{fig:PIC_profile} we present density and field profiles for a simulation with parameters $ \omega_{\mathrm{p}}= 0.27~\omega_{0}$, $R_{0}=81\,c/\omega_{0} $ and $ a_{0}=600 $ at time $ t = 2501~ \omega_{0}^{-1} $.
Panels (a) and (c) show density profiles of ions and electrons during the propagation of the laser pulse depicted in panel (e).
The detailed structure of the electromagnetic field within the smaller box marked in (e)  is presented in panels (b), (d) and (f).
Inside the channel, the laser is partially absorbed, but the wavefronts are nearly straight, as can be verified by observing the transverse electromagnetic field components $E_y$ and $B_z$. 
The longitudinal electric field $E_x$ generated by the density fluctuations within the channel is noisy, and can potentially work towards or against acceleration of individual particles. 
Panels (g) and (h) show the total focusing field associated with the plasma channel, given by equation \eqref{channel_fields_eq}.
As these fields are weaker than the laser field, one can obtain only the total value $|E_\mathrm{C}|+|B_\mathrm{C}|$ by subtracting $E_{y}-B_{z} $. Doing so reveals a long-range field structure, linearly dependent on the distance from the axis. 
The amplitude of the focusing field is of the same order as $E_x$, with a difference that the $E_x$ average is close to zero on scales larger than a laser wavelength.  
%
%
The color-coded lineouts in panel (h) confirm that the channel field can be regarded as a linear function of the distance from the axis up to approximately $ R_{0}/2 $, where the
slope is defined by the effective plasma density. 
For a flat density plasma slab, the effective density is the background plasma density $n_\mathrm{p}$. 
For a channel with a parabolic density gradient like ours, the average plasma density for the region around the channel axis between  $-R_{0}/2 $ and $ R_{0}/2 $ can be calculated analytically and directly applied to equations presented in Section I.
The channel field predicted by equation \eqref{channel_fields_eq} using the effective plasma density $n^\mathrm{eff}_\mathrm{p}\simeq 0.4~n_\mathrm{c}$ is illustrated by the black dashed line in panel (h) which is in agreement with the value extracted from the PIC simulation.
%

%
\subsection{Particle motion}

\begin{figure*}
		\centering
		\includegraphics[width=1\linewidth]{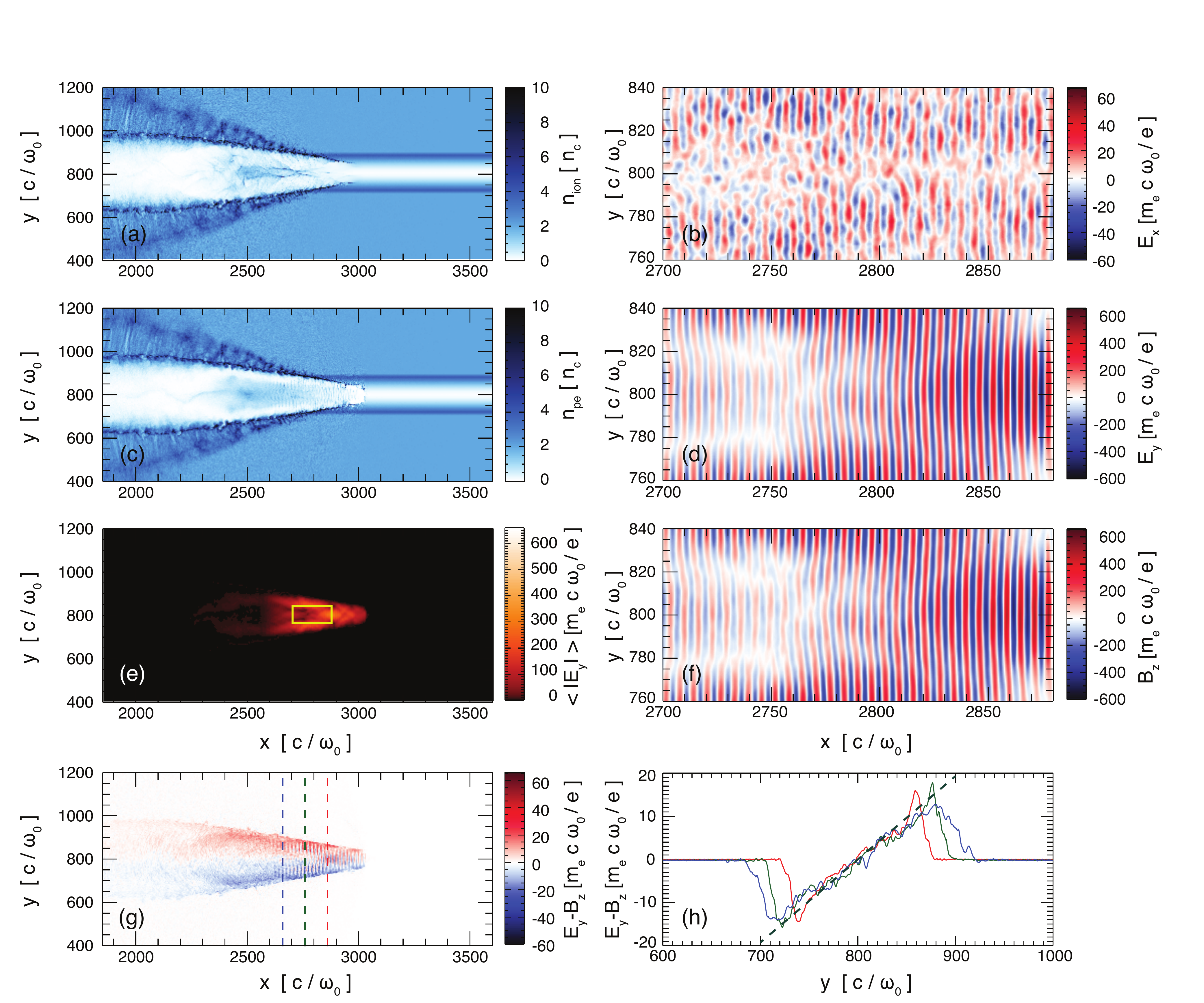}
		\caption{Results from PIC for $ 
					\omega_{\mathrm{p}}= 0.27~\omega_{0}$, $R_{0}= 
					81\,c/\omega_{0} $ and $ 
					a_{0}=600 $ at time $ t = 2501~ \omega_{0}^{-1}$ . (a) Proton density, (c) electron density, (e) 
					average absolute value of transverse electric field, (b) longitudinal electric field, (d) transverse electric field,  (f) transverse magnetic field, (g) background channel field obtained by subtracting $E_y-B_z$ and (h)  lineouts of the channel field. The black dashed line in panel (h) represents the expected channel field according to equation \eqref{channel_fields_eq} using  $n_\mathrm{p}^\mathrm{eff}\simeq0.4~n_\mathrm{c}$.
		}
		\label{fig:PIC_profile}
	\end{figure*}

\begin{figure*}
		\centering
		\includegraphics[width=1\linewidth]{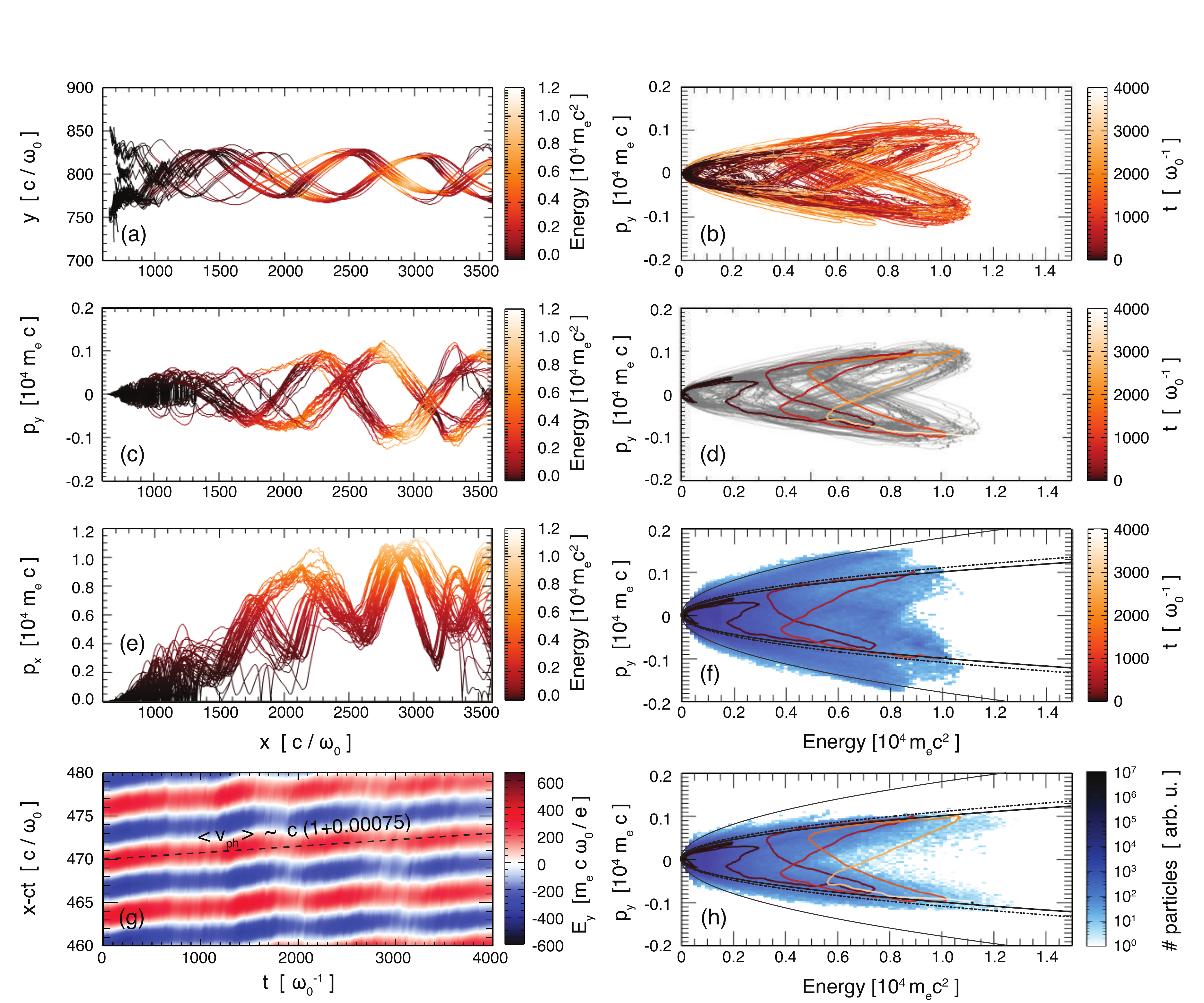}
		\caption{Results from PIC for $ 
					\omega_{\mathrm{p}}= 0.27~\omega_{0}$, $R_{0}= 
					81\,c/\omega_{0} $ and $ 
					a_{0}=600 $. (a) Trajectory, (c) transverse and (e) longitudinal momenta of 48 randomly selected electrons with energy above $10^4~m_e c^2$ at $ t = 2501~ \omega_{0}^{-1}$. Panels (b, d, f, h) show the time evolution of the electrons in $ p_{y} $-energy phase-space. Panel (d) highlights the trajectory of one efficiently accelerated electron. 
					This trajectory is overlaid with the $ p_{y} $-energy 
					phase-space (blue scale) and analytical estimates for the 
					connection between $p_y$ and energy when betatron phase is 
					$\cos\psi\simeq1$ (the point of local maximum for $p_y$) in 
					panels (f) $ t = 2501~ \omega_{0}^{-1}$ and (h) $ t = 4000~ 
					\omega_{0}^{-1} $. The outer solid line is associated with 
					the highest initial integral of motion in the system, while 
					the inner lines show the expected values after one (dotted 
					line) and two (solid line) resonant cycles according to 
					equation 
					\eqref{IOM_reduced}. Panel (g) shows time evolution of the 
					transverse electric field along the $ x $-axis inside a 
					simulation window moving at the speed of light. The figure 
					represents a magnified portion of space slightly over three 
					wavelengths long, such that phase velocity can be measured 
					directly.
		}
		\label{fig:PIC_trajectory}
	\end{figure*}

For the same simulation parameters, figure~\ref{fig:PIC_trajectory} shows randomly selected trajectories among particles that achieved energies $\xi>10^4~m_e c^2$ at $t=2501~ \omega_{0}^{-1}$ (the same time shown in figure \ref{fig:PIC_profile}).
The trajectories are color-coded in energy, showing the configuration space along with the evolution of transverse and longitudinal momenta vs. the longitudinal $ x $-position.
Laser-electron dephasing depends on the velocity of the particle along the laser propagation direction $\beta_x$ and phase velocity of the laser.  
One can measure the phase velocity in PIC simulations from the temporal evolution of the electric field along the channel axis, shown in panel (g).  
The phase velocity is not constant, owing to the 2D laser dynamics. 
The measured average phase velocity is $ 1.00075c $, on the same order as the analytical prediction in Section II that gives the value $ 1.00066c $ for our $n^{\mathrm{eff}}_\mathrm{p}\simeq 0.4~n_\mathrm{c}$ and $a_0=600$. 
The $\beta_x$ depends on the ratio $p_y/p_x\sim 0.1$ (c. f. figure~\ref{fig:PIC_trajectory}), which confirms that for resonant particles the transverse electron motion has about an order of magnitude stronger effect on dephasing than superluminosity  (and more than one order of magnitude for non-resonant particles). 
We can therefore apply our analytical model for studying the particle motion.
Figure \ref{fig:PIC_trajectory} shows the particles perform a similar asymptotic motion at late times, despite the different initial dynamics.
Some particles achieve resonance faster than others.
The interaction with the laser can start from rest, or after pre-acceleration at the laser-plasma interface.  
The integral of motion associated with the initial distance $R_0/2$ from the axis is $\mathcal{I}_0\sim160$.
According to equation \eqref{IOM_reduced}, the expected integral of motion $\mathcal{I}$ after one full resonant cycle would be $\mathcal{I}\sim60$, while after two cycles $\mathcal{I}\sim50$. 
Among our selected particles, there are many that completed two full resonant cycles. 
The temporal evolution of trajectories in $ p_{y} $-energy phase-space is also shown, with a highlighted single-particle example in panel (d).
This illustrates how the most energetic particles gain energy over time: 
they first oscillate in the region of low energy, unable to retain the energy gained over the oscillation cycle;  
as the oscillations become quasi-resonant, the particles can retain a fraction of the energy obtained, and they accelerate sloshing back and forth from the points on the parabola defined by their current integral of motion through $ p_{y}=m_{e}c\sqrt{2\gamma \mathcal{I}} $. The definition of the parabola follows from equation~\eqref{IOP_Wang} when $ \cos\psi\simeq1 $. 
The phasespace associated with simulation particles initially at rest within the channel between $680<x*\omega_0/c<725$ is shown in panels (f) and (h) for times $ t = 2501~ \omega_{0}^{-1}$ and $ t = 4000~ \omega_{0}^{-1}$, respectively.
The colorscale is logarithmic, to highlight all possible combinations of $p_y$ and energy.
The widest parabola is associated with the maximum initial  integral of motion $\mathcal{I}_0=160$ available among the considered particles.
As the time passes, the invariant $\mathcal{I}$ of some particles can be reduced.
The two inner parabolas show the asymptotic value after one (dashed line) and two (solid line) resonant cycles. 
The energetic particles are eventually expected inside the parabola associated with the asymptotic $\mathcal{I}$, as shown in panel (h). 
According to equation \eqref{gamma_max_channel0_cgs}, the expected asymptotic value of the Lorentz factor for $\mathcal{I}=50$ is $\gamma\sim 1.1\times10^4$.
We can also predict the amplitude of oscillations around the channel axis for these particles is about $22~c/\omega_0$ by inserting $p_y=0$ in equation \eqref{IOP_Wang}. 
All these estimates are in agreement with our simulation results. 
%
%
%
%
%
%
%

	\section{IV. Conclusions}
	
The betatron resonance in a plasma channel can result in efficient electron acceleration, but strongly depends on the initial particle position, channel density, laser intensity and acceleration distance.
	Particles with an initial radius $R_0\simeq R_\mathrm{opt}$, where $R_\mathrm{opt}$ is defined by equation \eqref{optimal_r}, have the best chance to attain the maximum energy in the system. 
	At very high laser intensities, radiation reaction causes a decrease in the 
	maximum electron energy, however, it allows accelerating a higher number of 
	particles as it affects the phase-matching process between betatron and laser 
	oscillations.
	In fact, nearly all particles with  $R_0> R_\mathrm{opt}$ have a very good chance to be accelerated due to radiation reaction, provided that 
	 $R_\mathrm{opt}$ is still within the range where channel fields are nearly a linear function of the distance from the laser propagation axis.  
	To that end, it may be optimal to use plasma channel radius $\sim 2R_\mathrm{opt}$.

	Analytical predictions of the maximum electron energy as a function of channel 
	density and electron initial radial position are provided by equations \eqref{gamma_final_max} and \eqref{gamma_max_channel0} using radiation-reaction reduced value of $\mathcal{I}$ given by the lowest estimate from equations \eqref{IOM_reduced} and \eqref{IOM_reduced_laser}. 
	The required acceleration distance to achieve this energy can be estimated through equation \eqref{gamma_l}.
	Our findings are confirmed by numerical integration of test particle trajectories, as well as 2D full-scale PIC simulations. 
	According to the presented calculations, using near-future lasers ($ 10~\mathrm{PW} $-class, $ 150~\mathrm{fs} $) we can 
	obtain multi-GeV electrons in a single-stage acceleration within a 0.5 mm-long plasma channel. 
	We envisage a roadmap towards electron energies in excess of $10~\mathrm{GeV}$ if the channel and laser parameters are matched for an optimal outcome. 

	Our findings can be extended to acceleration of externally injected beams, 
	with an arbitrary beam loading (i.e. high current within the channel, 
	similar as in reference \cite{Gong2019}). In that case, the background 
	magnetic field due to the current within the channel may be stronger than 
	expected from the background plasma density. For such a setup, one should 
	estimate an effective (higher) value for the plasma density 
	$n_\mathrm{p}^\mathrm{eff}$ that would correspond to a comparable current 
	density and the same order of magnitude for the self-consistent background 
	magnetic field within the plasma. Our scaling laws can then be applied 
	using this effective $n_\mathrm{p}^\mathrm{eff}$ instead of $n_\mathrm{p}$ 
	to calculate the asymptotic cutoff energy of the electron beam. An example 
	for $a_0=200$ and $a_{MB}=10
^{-2}a_0$ from reference \cite{Gong2019}, can be mapped using 
$\mathcal{I}\sim40$ and $\omega_{\mathrm{p}}/\omega_0\sim0.3$. Our model 
predicsts the 
asymptotic electron energy is $\sim$~7~GeV, which is in agreement with their 
simulations that show 7.5 GeV. 
	
	\section{Acknowledgements}
	This work was supported by the European Research Council (ERC-2015-AdG Grant No. 695088), Portuguese Science Foundation (FCT) Grant No. SFRH/BPD/119642/2016 and by the project 
	High Field Initiative (CZ.02.1.01/0.0/0.0/15\_003/0000449) from European 
	Regional Development Fund (HiFI).
	The support of Czech Science Foundation project No. 18-09560S is acknowledged.
	The results of the  Project LQ1606 were obtained with the financial support of 
	the Ministry of Education, Youth and Sports as part of targeted support from 
	the National Programme of Sustainability  II.
	We acknowledge PRACE for awarding access to MareNostrum based in the Barcelona Supercomputing Center. 
	The simulations were performed at MareNostrum (Spain), IST Cluster (Portugal) 
	and ECLIPSE (ELI Beamlines, Czech Republic).
	%
	%
	
	
	%
	
	%
	
	%
	
	%
	
	%
	
	%
	
	%
	
	%
	
	%
	

	
	
	
	\bibliography{channel}{}
	\bibliographystyle{iopart-num}

\end{document}